\documentclass[11pt,british]{aa}
\usepackage{graphicx}
\usepackage{txfonts}
\usepackage{amsfonts}

\usepackage{url}       
\usepackage[colorlinks=true,citecolor=blue]{hyperref}

\makeatletter
\newcommand{\arcdeg}{$^\mathrm{o}$}
\newcommand{\ringpatch}{$16^\mathrm{o}\times16^\mathrm{o}$}

\newcommand{\an}{Astron. Nachr.}
\newcommand{\arfm}{   {\it Ann. Rev. Fluid Mech.}}
\newcommand{\aspcs}{Astron. Soc. Pac. Conf. Ser.}
\newcommand{\asr}{    {\it Adv. Spa. Res.}}
\newcommand{\basi}{Bul. Astron. Soc. India}
\newcommand{\jaa}{    {\it J. Astrophys. Astron.}}
\newcommand{\jpcs}{   {\it J. Phys. Conf. Ser.}}
\newcommand{\vag}{Vierteljahrsschr. Astron. Ges.}

\makeatother

\begin{document}
\title{Activity Related Variations of High-Degree p-Mode Amplitude, Width and Energy in Solar Active Regions}
\titlerunning{Variations of p-Mode Amplitude, Width and Energy in Active Regions}

\author{R. A. Maurya\inst{1,2}, A. Ambastha\inst{2} and J. Chae\inst{1}}
\authorrunning{R.A. Maurya, et al.}

\institute{Department of Physics and Astronomy, Seoul National University, Seoul 151-747, Republic of Korea.\\
\email{ramajor@astro.snu.ac.kr, jcchae@astro.snu.ac.kr}\and Udaipur
Solar Observatory, Physical Research Laboratory, Udaipur-313001, India\\
\email{ambastha@prl.res.in}}

\offprints{R. A. Maurya}

\date{Received: 6 February 2013; Accepted...}

\abstract
{Solar energetic transients, e.g., flares, CMEs, etc., occurring mostly within active regions (ARs), release large amount of energy which is expected to excite acoustic waves by exerting mechanical impulse of the thermal expansion of the flare on the photosphere. On the other hand, strong magnetic fields of sunspots of ARs absorb the photospheric oscillation modes' power.}
{We study the properties of high degree p-mode oscillations in flaring and dormant ARs and compare them with those in corresponding quiet regions (QRs) to find the association of mode parameters with magnetic and flare related activities.}
{We computed the mode parameters using the ring-diagram technique. The magnetic activity indices (MAIs) of ARs and QRs are determined from the line-of-sight magnetograms. The flare indices (FIs) of ARs are obtained from the GOES X-ray fluxes. Mode parameters are corrected for foreshortening, filling factor and MAI, using multiple non-linear regression.}
{Our analysis of several flaring and dormant ARs observed during the Carrington Rotations 1980-2109, showed strong association of mode amplitude, width and energy with magnetic and flare activities although their changes are combined effects of foreshortening, filling factor, magnetic activity, flare activity, and measurement uncertainties. We find that the largest decrease in mode amplitude and background power of an AR are caused by the angular distance of the AR from the solar disc centre. After correcting the mode parameters for foreshortening and filling factor, we find that the mode amplitude of flaring and dormant ARs are smaller than in corresponding QRs, and decreases with increasing MAI suggesting a larger mode power suppression in ARs with stronger magnetic fields. The mode widths in ARs are larger than in corresponding QRs and increase with MAI, indicating shorter lifetimes of modes in ARs than in QRs. The variations in  mode amplitude and width with MAI are not same in different frequency bands. The largest decrease (increase) in mode amplitude (mode width) of dormant ARs is found in the five minute frequency band. The average mode energy of both the flaring and dormant ARs are smaller than in their corresponding QRs,  decreasing with increasing MAI. But the average mode energy decrease rate in flaring ARs are smaller than in dormant ARs. Also, the increase in mode width rate in dormant (flaring) ARs is followed by decrease (increase) in amplitude variation rate. Furthermore, inclusion of mode corrections for MAI shows that mode amplitude and mode energy of flaring ARs increase with FI while mode width shows an opposite trend. The variations in mode parameters with FI are also not the same with frequency. The increase (decrease) in mode amplitude (width) is larger in the five minute and higher frequency band. The increase in width variation rate is followed by rapid decrease in the amplitude variation rate.}{}


\maketitle

\section{Introduction}

Photospheric five-minute oscillations, probably first observed by \citet{Leighton1962}, are caused by trapped acoustic waves (p-modes) inside the solar interior \citep{Ulrich1970, Leibacher1971}, are well known and have been studied extensively. It is believed that the energy of p-modes is contributed by convective or radiative fluxes. Precise determination of the properties of p-modes provides a powerful tool to probe the solar interior. High-degree ($\ell>200$) acoustic oscillations are vertically trapped in a spherical shell with the photosphere as the upper boundary and the lower boundary depending on the horizontal wavenumber, $k_h^2=\frac{\ell(\ell+1)}{r^2}$, and the frequency ($\omega$),

\begin{equation}
\frac{\ell(\ell+1)}{r_t^2}=\frac{\omega^2}{c_s^2(r_t)}
\label{eq:lower-turn-pt}
\end{equation}

\noindent where, $r_t$ is depth of the lower turning point. Lifetimes of high degree modes are much smaller than the sound travel time around the Sun, therefore local effects are more important for these modes than for the low degree modes which have larger horizontal wavelengths and longer lifetimes. It is likely that high degree acoustic waves are not global modes, that is, they do not remain coherent while travelling over the circumference to interfere with themselves. So locally they can be considered as horizontally travelling, vertically trapped waves. These are observed as photospheric motions inferred from the Doppler shifts of photospheric spectral lines. The analysis of local modes provides a diagnostic tool to study the structural and dynamic properties of solar interior beneath ARs, as well as, the changes in excitation and damping of modes that are most affected by the surface magnetic fields.

Earlier studies showed that characteristics of the high-degree modes are modified in ARs possessing complex, and strong magnetic fields associated with sunspots as compared to magnetically quiet regions. Significant reduction of p-mode energy in ARs has been attributed to absorption by sunspots \citep{Braun1987, Braun1990, Bogdan1993, Zhang1997, Hindman1998, Haber1999a, Mathew2008, Gosain2011}. Their studies showed that mode power absorption changes with harmonic degree and radial order. The absorption increases with increasing horizontal wavenumber $k_{\rm h}$ over the range 0.0\,--\,0.8\,Mm$^{-1}$ and decreases for larger $k_{\rm h}$, in the range 0.8\,--\,1.5\,Mm$^{-1}$. The absorption along each individual p-mode ridge tends to peak at an intermediate value of the degree 200\,--\,400. \citet{Hindman1998} found that the amplitudes of oscillations with frequencies less than 5.2 mHz decrease with field strength for both velocity and continuum intensity measurements. Furthermore, they reported that oscillations with frequencies between 5.2 and 7.0 mHz within ARs suppressed continuum intensity amplitudes but enhanced velocity amplitudes.

The differences in mode frequency and width in ARs and QRs have also been studied earlier. \citet{Rajaguru2001} found that the frequencies of solar oscillations are significantly higher in the ARs than in QRs. Width and asymmetry of the peaks in power spectra are also larger in ARs while the mode amplitudes are smaller. Furthermore, \citet{Howe2004} showed that the difference in mode characteristics correlated with the differences in the average surface magnetic fields between corresponding regions. They reported strong dependence of amplitude and lifetime of p-modes on the local magnetic flux, and found that these parameters decreased in the 5 minute band while a reverse trend was found at high frequencies. \citet{Rabello-Soares2008} confirmed the earlier results and reported that the mode amplitude and width variations are nearly linear.

Changes in the mode parameters with solar activity cycle have also been studied by many researchers. \citet{Tripathy2010} found that changes in mode frequency during the activity minimum period are significantly greater than in the solar maximum. \citet{Komm2000} reported a 23\% increase in mode widths (i.e., decrease in mode lifetimes) with increasing solar activity. They also found that the variations are frequency dependent; being largest near 3.1\,mHz, and independent of harmonic degree $\ell$. Changes in mode amplitudes, power, and energy-supply rate were also analysed from solar minimum to maximum. It was found that mode amplitudes and solar activity level are anti-correlated for intermediate and high-degree modes, and strongly depended on local magnetic fields. \citet{Chaplin2000} suggested that the changes in mode parameters arise from changes in damping rather than excitation. \citet{Burtseva2009} found that the mode lifetime in ARs and QRs decrease with magnetic activity, but the decrease is relatively slower in QRs. Also, p-mode amplitudes at solar minimum are higher than at solar maximum \citep{Burtseva2009a}.

Energetic transients, viz., flares, CMEs, etc., are believed to occur due to the reconnection of magnetic fieldlines in the solar atmosphere. Particles are energized at the primary energy release site in the corona and then guided along the magnetic fieldlines downwards to the denser atmosphere. They are expected to affect and excite the high degree p-modes by exerting mechanical impulse of thermal expansion on the photospheric layer \citep{Wolff1972}. The observational evidence of  flare induced horizontally travelling waves on the solar photosphere have been reported earlier \citep{Kosovichev1998, Donea1999, Donea2005, Kosovichev2006a, Kosovichev2011a}. The detection of seismic waves provides us with unique opportunity to study the excitation of solar oscillations, and raise new questions about the underlying physical processes as well as the properties of the excited waves and source producing them. But their observations are relatively rare, possibly due to the difficulties of detecting the photospheric ripples. Such travelling waves appear to be associated mostly with energetic flares.

Flare related variations in high-$\ell$ p-mode parameters have been studied earlier \citep{Ambastha2003a, Maurya2009f, Maurya2010, Maurya2011a}. \citet{Ambastha2003a} have found that power in high-degree p-modes is amplified during periods of high flare activity in some ARs than in the non-flaring ARs of similar magnetic fields. \citet{Maurya2009f} found strong evidence of substantial increase in mode amplitude during an extremely energetic flare. On the other hand, at global scale of the whole solar disk, \citet{Ambastha2006a} found a poor correlation between the running mean of FI and low-$\ell$ mode power. They reported similar changes for CME index. \citet{Karoff2008} found that energy in the flare can drive global oscillations in the Sun in the same way that the Earth is set ringing after a major earthquake. They reported a larger correlation between flares and energy of oscillations at high frequencies than for the ordinary p-modes. Flare induced excitation at higher frequencies are also reported by \citet{Kumar2010, Kumar2011}. A more recent study by \citet{Richardson2012}, however, did not find evidence of flare-driven high frequency global modes. It may be noted that these high frequency modes, unlike the p-modes, are not expected to be trapped in the solar interior.

Solar ARs, where  the flares usually occur, are locations of strong magnetic fields associated with the sunspots. Thus, there is a competition between absorption of p-mode energy by sunspots and excitation/amplification by the energetic transients. The net increase or decrease of mode energy would, therefore, depend on their relative dominance. Characteristics of the photospheric oscillations in ARs are essentially described by the modes of shorter wavelength that are trapped just below the photosphere. These modes can be studied using the so-called ring diagram analysis \citep{Hill1988}, which is a well-established technique in local helioseismology. It is a generalization of the global-mode analysis for small patches of the solar surface, where the geometry is approximately plane-parallel, and waves can be treated as plane waves. The modes measured using this technique are typically $\ell>150$, which have lower turning depth above $r/R_\sun\approx0.95$. For more detail about this technique, readers may refer to the review by \citet{Antia2007}.

The ring-diagram technique uses several hours' long time-series of Doppler observations which may lead to dilute the rather shorter duration flare-induced effects owing to the averaging involved in space and time. Nevertheless, we found strong evidence of flare related amplification in p-mode amplitude during a long duration, large magnitude X17.2/4B flare of 28 October 2003 \citep{Maurya2009f}. This study illustrated the possibility of detection of changes in p-mode parameters for long duration energetic flares by ring-diagram analysis.  The observed amplification in p-mode amplitude implies that during an energetic flare, the energy of p-modes must be weighted by the energy of excited modes. In order to further establish these results, it is required to investigate several ARs, both flare productive and dormant along with QRs. In this paper, we undertake such a study involving a set of 53 flaring ARs of varying magnetic complexity to examine the relationship of mode amplitude, width and energy with magnetic fields and flare energies, however, ring-diagram analysis can also be used to study frequency shift, sub-photospheric flow, sound speed, etc.

\begin{figure*}[ht]
\centering
\includegraphics[width=0.8\textwidth,bb=15 3 429 276]{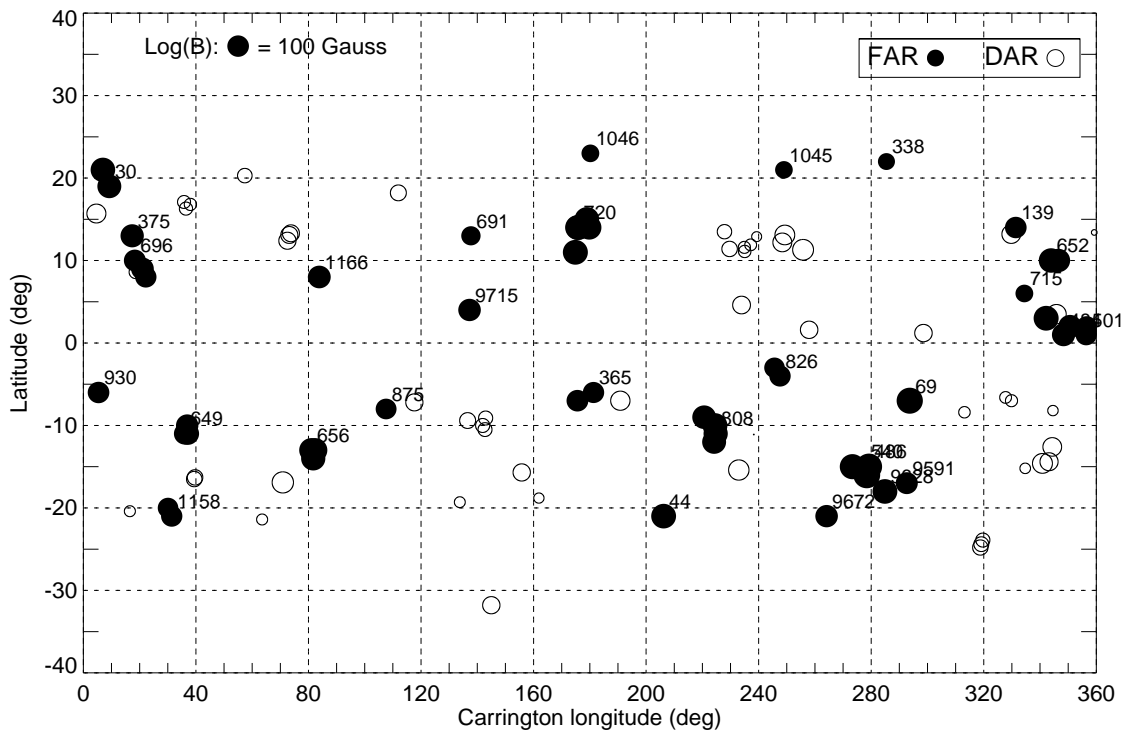}
\caption{Distribution of flaring (filled circle) and dormant (unfilled circle) ARs used in this study as shown in a Carrington map. Circle sizes correspond to the magnetic activity index (B; for detail see Section~\ref{subsec:MAI-FI}
) of ARs in Gauss.}\label{fig:ars_distbsn}
\end{figure*}

The paper is organized as follow: Section \ref{sec:ARsData} describes the selected sample of flaring and dormant ARs and the observational data used for our study. The methods of data analysis and results are given in Section~\ref{sec:Analys}. We briefly discuss how the mode parameters are affected by foreshortening, filling factors, and magnetic activities, and present an empirical method to correct them. We assume that these effects are the only systematic variations and that measurement uncertainties are essentially random and are compensated for by averaging over many data sets. In Section~\ref{sec:ResultsDiscs} we present results of the analysis and their discussions.  We study variations in mode parameters with magnetic and flare activity indices of ARs. Note that, in this paper, we present average properties of mode parameters rather than their temporal variations. Section \ref{sec:SumConclsn} gives the summary and conclusions.

\section{Active Regions and Observational Data}
\label{sec:ARsData}

\subsection{Flaring and Dormant Active Regions}
\label{subsec:FlareDormARs}

We selected our sample of energetic events occurring in different ARs using the archived information on ARs and solar activity from the web-pages of the Solar Monitor~\footnote{\url{http://www.solarmonitor.org}}. We first identified flares of X-ray classes $>$ M5.0 during the period of Carrington rotations 1980--2109 of solar cycles 23 and 24. We short-listed the ARs lying within the angular distance of 40\arcdeg~from the solar disc centre and selected 53 events that were well covered by the Global Oscillation Network Group (GONG) project. Out of 53 events, 48 events correspond to solar cycle 23 and 5 from the ongoing solar cycle 24. We also selected suitable quiet regions (QRs) corresponding to every flaring AR located at the same latitude but at a different longitude.

In order to study the flare induced excitation in flaring ARs as compared with non-flaring ARs, we also selected dormant ARs and associated QRs as for the flaring ARs. Thus for every flare event, we selected four regions: flaring and dormant ARs and their associated QRs, for the same time period. The locations and distribution of flare events and corresponding ARs over the solar disc are illustrated in Figure~\ref{fig:ars_distbsn}. Here onwards, for brevity, we have used the notation B for magnetic activity index (MAI) of an AR, measured in Gauss, which is not to be confused with magnetic field strength. Most of the events selected in our sample occurred within the latitude zone of $\pm20^{\rm o}$, therefore, projection effects are not expected to be important for these ARs.

\subsection{The Observational Data}
\label{sec:data}

The observational data used in this study consist of merged Dopplergrams at one minute cadence, obtained by the GONG \citep{Harvey1988}, 96-minutes averaged magnetograms from Michelson Doppler Imager \citep[MDI,][]{Scherrer1995} on board Solar and Heliospheric Observatory (SOHO), and the X-ray fluxes from the GOES data archive. The pixel resolutions for GONG Dopplergrams and MDI magnetograms are 2.5\arcsec~and 2.0\arcsec, respectively. The 96-minutes averaged magnetograms provided by MDI data archive are averages of calibrated one-minute magnetograms (with noise level of $\sim30$\,G) and five-minute magnetograms (with noise level of $\sim15$\,G).

\section{The Analysis Techniques}
\label{sec:Analys}

\subsection{Ring Diagrams and p-mode Parameters}
\label{subsec:ring-analysis-mode-par}

In order to estimate the p-mode parameters corresponding to a selected area over the Sun, the region of interest is tracked over time. Such a spatio-temporal area is defined by an array (or data-cube) of dimension $N_x\times N_y \times N_t$. Here, first two dimension ($N_x, N_y$) correspond to the spatial size of the AR along $x$- and $y$-axes, representing zonal and meridional directions, and the third ($N_t$) to the time $t$ in minutes. The data-cubes employed by ring diagram analysis have typically 1664 minutes' duration and cover $16^{\rm o}\times16^{\rm o}$~ area centred at the location of interest. This choice of area makes a compromise between the spatial resolution on the Sun, range of depths and resolution in spatial wavenumber of the power spectra. A larger size allows access to the deeper sub-photospheric layers, but only with a coarser spatial resolution. On the other hand, a smaller size not only limits access to the deeper layers but also renders the fitting of rings more difficult.

The spatial coordinates of pixels in tracked images are not always integer. To apply the three-dimensional Fourier transform on tracked data cube, we interpolate the coordinates of tracked images to integer values, for which we use ``sinc'' interpolation method. Three-dimensional Fourier transformation of data cube produces truncation of rings near the edges due to the aliasing of higher frequencies toward lower side. In order to avoid the truncation effects, we apodize the data cube in both the spatial and temporal dimensions. The spatial apodization is obtained by a 2D-cosine bell method which reduces the $16^{\rm o}\times16^{\rm o}$ area to a circular patch having a radius of $15^{\rm o}$ \citep{Corbard2003}.

The observed photospheric velocity signal $v(x,y,t)$ in the data cube is a function of position $(x, y)$ and time $(t)$. Let the velocity signal in frequency domain be $f(k_x, k_y,\omega)$, where, $k_{\rm x}$~and $k_{\rm y}$~are spatial frequencies in $x$- and $y$- directions, respectively, and $\omega$ the angular frequency of oscillations. Then the data cube $v(x,y,t)$ can be written as,

\begin{equation}
v(x,y,t) = \int {\int {\int {f(k_x ,k_y ,\omega)} } } e^{i(k_x x + k_y y + \omega t)}\,dk_x\,dk_y\,d\omega.
 \label{eq:VelCube}
\end{equation}

\noindent The amplitude $f(k_x ,k_y ,\omega)$ of p-mode oscillations is calculated using three-dimensional Fourier transformation of Equation [\ref{eq:VelCube}]. The power spectrum is given by,

\begin{equation}
P(k_x, k_y, \omega) = |f(k_x ,k_y ,\omega)|^2.
 \label{eq:PowCube}
\end{equation}

The 16$^{\rm o}$~patch from GONG consists of 128 pixels, giving a spatial resolution $\Delta x =$1.5184\,Mm, i.e., the $k$-number resolution, $\Delta k=3.2328\times10^{-2}$\,Mm$^{-1}$, and Nyquist value and resolution for the harmonic degree ($\ell$), approximately 1440 and  22.5, respectively. The corresponding range in $k_{\rm x}$-$k_{\rm y}$~space is~ (-2.069, 2.069)\,Mm$^{-1}$. The temporal cadence and duration of data-sets give Nyquist frequency of 8333\,$\mu$Hz~and frequency resolution of 10\,$\mu$Hz, respectively.

In order to determine various p-mode parameters, the three-dimensional power spectrum is fitted using the Lorentzian function \citep{Haber2002, Hill2003} given by

\begin{equation}
P(k_{x},k_{y},\omega)=\frac{A^\prime}{(\omega-\omega_{0}+k_{x}U_{x}+k_{y}U_{y})^{2}+\Gamma^{2}}+\frac{b_{0}}{k^{3}}
\label{eq:mode-fitting}
\end{equation}

\noindent where, $k=\sqrt(k_{x}^2+k_{y}^2)$ is the total horizontal wavenumber. It can be identified with the degree ($\ell$) of a spherical harmonic mode of global oscillations by $\ell(\ell+1)=k^2R^2_\sun$. The six fitting parameters, viz., zonal velocity ($U_{{\rm x}}$),  meridional velocity ($U_{{\rm y}}$), background power ($b_{0}$), mode's central frequency ($\omega_{0}$), mode width ($\Gamma$) and the parameter $A^\prime$ are determined by the maximum-likelihood approach \citep{Anderson1990}. Note that $A^\prime$ is not amplitude (or mode height); the amplitude (say, $A$) can be determined from the expression, $A^\prime=A\times\Gamma^2$. For our convenience we use the term ``mode'' to refer these parameters although these are not discrete modes in the sense as those observed at lower degrees. In this paper, we are mainly interested in the flare and magnetic activity associated changes in mode parameters: $A$, $b_0$, $\Gamma$ and mode energy (see Section~\ref{sec:mode-en}).

\begin{figure*}[ht]
\centering
\includegraphics[width=0.8\textwidth, bb=31 6 408 456]{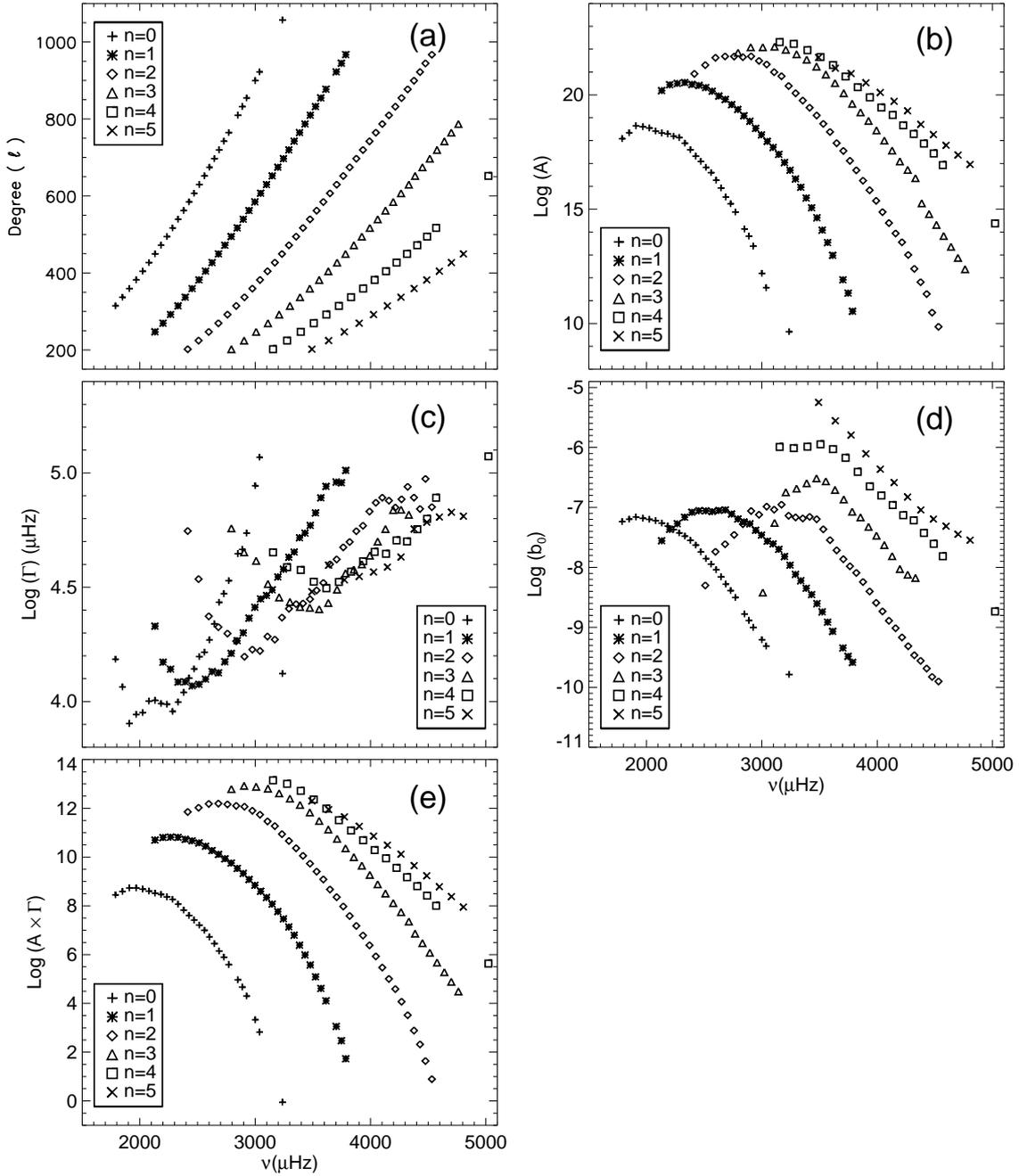}
\caption{The p-mode parameters obtained for AR NOAA 10649 on 17 July 2004: (a) harmonic degree $\ell$, (b) mode amplitude $A$,  (c) mode width $\Gamma$, (d) background power $b_0$, and (e) mode area $A\times\Gamma$, as a function of frequency for radial orders $n=0,\ldots,5$.}
\label{fig:ModPar}
\end{figure*}

A sample of p-mode parameters computed using the above method are shown in Figure~\ref{fig:ModPar} for AR NOAA 10649 obtained for the data cube on 17 July 2004. Panel (a) shows the well known $\ell-\nu$ diagram which represents the dispersion relation of the acoustic waves trapped in the sub-photospheric depths of the AR. Different ridges correspond to modes of different radial orders,  $n=0,\ldots,5$ . Modes with larger harmonic degree $\ell$ correspond to shorter wavelengths, which are trapped in shallower depths (see Equation~\ref{eq:lower-turn-pt}). These are the modes which are more likely to be modified by the near surface activities, such as, flares and sunspots' magnetic fields. Other panels show the natural logarithm values of the mode parameters, $A'$, mode width ($\Gamma$), and background power ($b_0$) as a function of frequency.

\subsection{The p-mode Energy}
 \label{sec:mode-en}

Photospheric p-mode parameters provide a diagnostic tool to study sub-photospheric structures and dynamics of ARs. For instance, mode width and amplitude provide clues to the excitation and damping mechanisms of solar oscillations. Line width ($\Gamma$) is directly related with lifetime, $(2\pi\Gamma)^{-1}$, of p-modes, and amplitudes can be converted into power and energy per mode. Total energy (i.e., kinetic and potential energy) of the p-mode can be given by \citep{Goldreich1994},

\begin{equation}
E_{n \ell}=M_{n \ell}<v_{n \ell}^{2}>
\label{eq:mode-energy}
\end{equation}

\noindent where, the parameter $M_{n.\ell}$ represents the mode mass which can be computed using the mode inertia, and $<v_{n \ell}^{2}>$ is mean square velocity, the measure of mode area. It is given by,

\begin{equation}
<v_{n \ell}^{2}>=\frac{\pi}{2} A_{n \ell} \Gamma_{n \ell}
\label{eq:mean-squareq:DataCube}
\end{equation}

\noindent where, $A_{n \ell}$ and $\Gamma_{n \ell}$ are mode amplitude and mode width, respectively, obtained from the ring fitting (see Equation~\ref{eq:mode-fitting}). The above expression for p-mode energy (Equation \ref{eq:mode-energy}) has been used  earlier for the study of solar cyclic variation of p-mode energy of global-modes, e.g.,  \citet{Komm2000a}. This expression for the mean square velocity is different from the expression given in \citet{Komm2000a}, where they used the correction factor $C_{\rm vis}=3.33$ for the reduced visibility due to leakage \citep{Hill1998}. But for the local modes, $C_{\rm vis}$=1. It is also to note that the mode widths computed from  ring-diagram analysis are larger than the actual values and hence the estimated mode energy will be large. Furthermore, there are many systematic effects on p-mode parameters obtained from ring diagram analysis that make it difficult to determine the absolute mode parameters of a given region. Most of these systematic effects can be eliminated, however, by studying the differences in the fitted parameters between different regions with same observing geometry. Therefore, we analyse mode parameters in ARs and corresponding QRs at same latitude but at different longitude. The relative value of mode energy, using Equations~\ref{eq:mode-energy} and~\ref{eq:mean-squareq:DataCube}, between an AR and corresponding QR is given by,

\begin{equation}
\frac{\delta E}{E}=\frac{(E_{n \ell})_{\rm AR}-(E_{n \ell})_{\rm QR}}{(E_{n \ell})_{\rm QR}} = \frac{(A_{n \ell}\Gamma_{n \ell})_{\rm AR}-(A_{n \ell}\Gamma_{n \ell})_{\rm QR}}{(A_{n \ell}\Gamma_{n \ell})_{\rm QR}}
\label{eq:ModeEn-ratio}
\end{equation}

This expression for the mode energy is free from mode mass, and hence we do not require mode inertia. Similarly,  we analyse fractional differences for other parameters, i.e., mode amplitude ($\delta A/A$), mode width ($\delta\Gamma/\Gamma$) and background power ($\delta b_0/b_0$) of flaring and dormant ARs along with their corresponding QRs.

\subsection{Magnetic and Flare Activity Indices}
\label{subsec:MAI-FI}

Magnetic activity index (denoted as $B$) for every data set of ARs and QRs is calculated from the 96-minutes averaged MDI magnetograms. Every magnetogram is tracked and remapped in the same manner as the Dopplergrams for ring-diagram analysis. Then areas (\ringpatch) of ring-patches are extracted from the full disc images. Then, magnetic activity index, $B$, is computed by taking absolute average over the patches. Since the MDI 96-minutes averaged magnetograms are obtained from both one- and five-minutes magnetograms, we set zero for those pixels which have values less than 35\,G. Mathematically, the MAI of an AR can be written as,

\begin{equation}
B=\frac{1}{N_2}\sum_{i=1}^{N_2}\left(\frac{1}{N_1}\sum_{|b|>b_{1}}|b|\right)
\label{eq:mai}
\end{equation}

\noindent where, $b_1$ is minimum threshold ($=35$\,G), $N_1$ is the number of pixels with $|b|> b_1$, $N_2$ is the number of mangetograms observed during the ring-diagram data cube. Thus, the values of $B$ for our samples of flaring and dormant ARs varies in the ranges of 37--308\,G and 15--90\,G, respectively, while for the QRs it ranges between 1\,--\,12\,G.

\begin{figure}[ht]
    \centering
        \includegraphics[width=0.50\textwidth,clip=,bb=20 6 354 273]{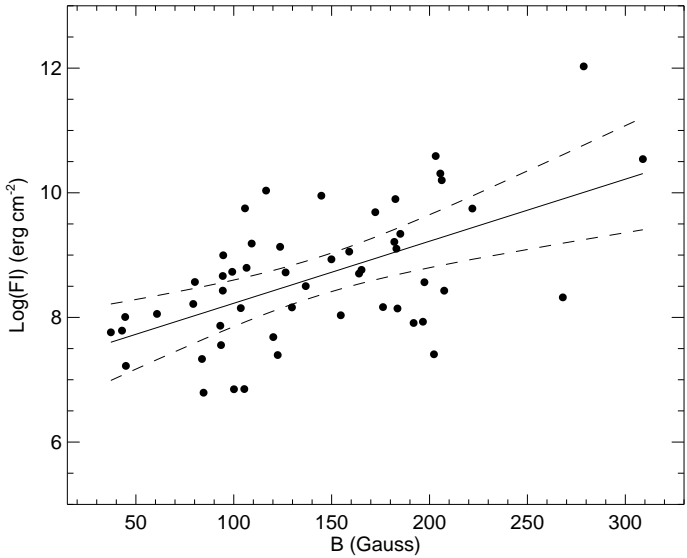}
    \caption{Relation between magnetic activity index ($B$) and flare activity index (FI) of  flaring ARs. Filled circle represent flaring ARs of our sample. The solid line shows the linear regression line while dashed lines around it correspond to 95\% confidence levels of linear fit.}
    \label{fig:mai_gflx}
\end{figure}

A measure of the flare activity of an AR, the X-ray flare index (say, FI), is calculated by multiplying the GOES X-ray flux with the flare duration, and then summing the contributions of all the flares that occurred during the 1664 minutes of the data-cube for the given AR. The flare index of an AR thus can be written as,

\begin{equation}
{\rm FI} = \sum_{i=1}^{N}F_i\delta t_i
\label{eq:flare-en}
\end{equation}

\noindent where, $F_i$ is the GOES X-ray flux for $i^{\rm th}$ flare event, $\delta t_i$ is the time duration of the same event, and $N$ is the total number of flare events that occurred in the time period of the data cube. GOES X-ray flux for different flare classes are listed in Table~\ref{tab:goes-flux}. Note that here we have only considered the flares of C, M and X classes.

\begin{table}[ht]
\centering
\caption{GOES X-class flares and associated energy fluxes}
\begin{tabular}{|l|c|}
\hline
X-class & GOES X-ray Flux\\
        & $F$ (${\rm erg\,{\rm s}^{-1}\,cm}^{-2}$)\\
\hline
Xn   &  n$\times10^{-1}$    \\
Mn   &  n$\times10^{-2}$    \\
Cn   &  n$\times10^{-3}$    \\
Bn   &  n$\times10^{-4}$    \\
An   &  n$\times10^{-5}$    \\
\hline
\end{tabular}
\label{tab:goes-flux}
\end{table}

Thus the flare activity index computed for the flaring ARs is found to range from $8.9\times10^2$ for moderately active ARs to $1.6\times10^5$ erg\,cm$^{-2}$ for very productive ARs. Figure~\ref{fig:mai_gflx} shows the relation between the magnetic activity and flare activity indices of our sample of flaring ARs. The straight line fit through the data points shows that flaring activity index increases with magnetic activity index of ARs. The 95\% confidence levels (dotted curves) of the fitting further validate this linear relation between FI and $B$.

\begin{figure*}[ht]
    \centering
        \includegraphics[width=0.49\textwidth,clip=,bb=13 6 282 419]{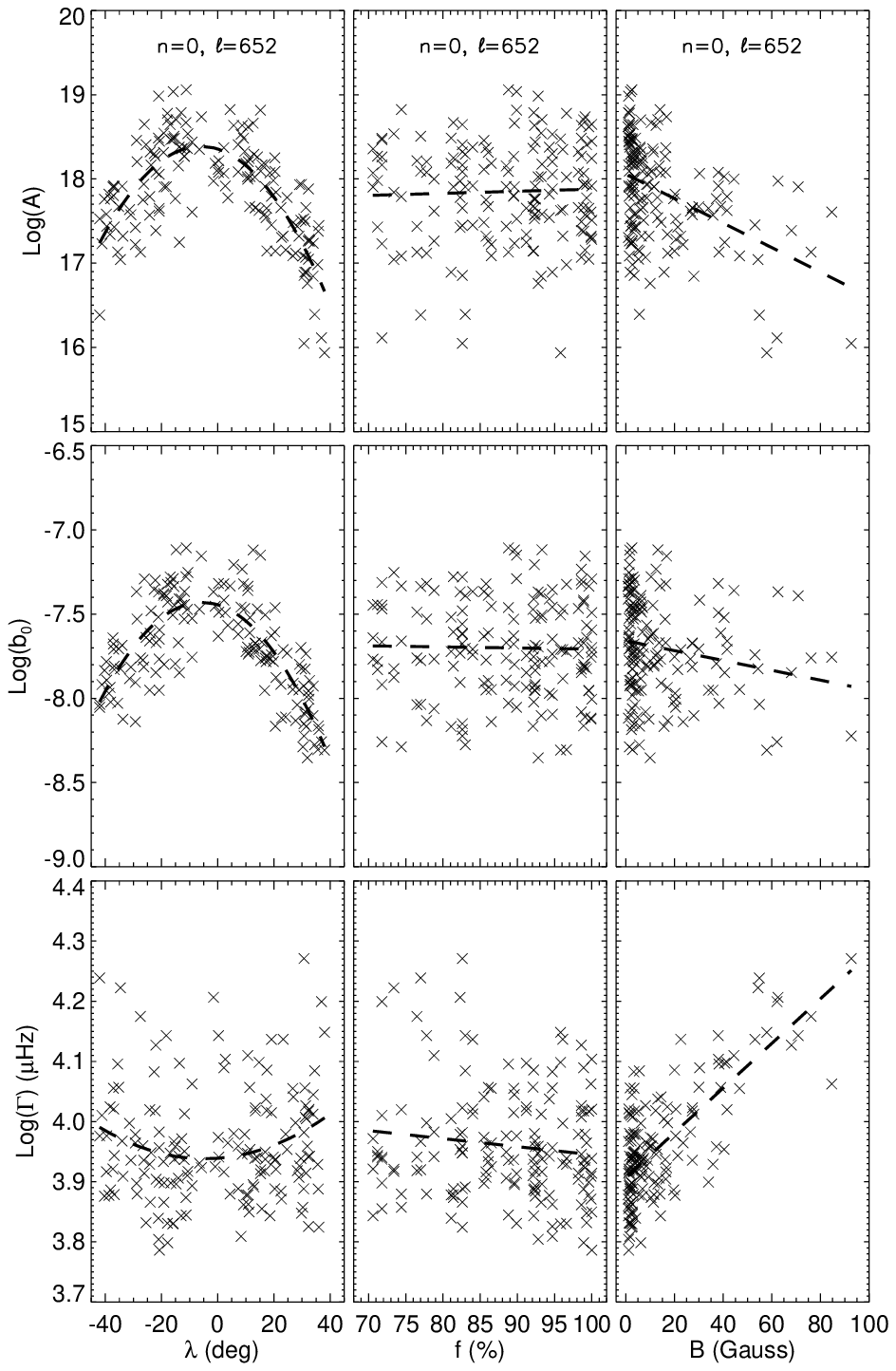}
        \includegraphics[width=0.49\textwidth,clip=,bb=13 6 282 419]{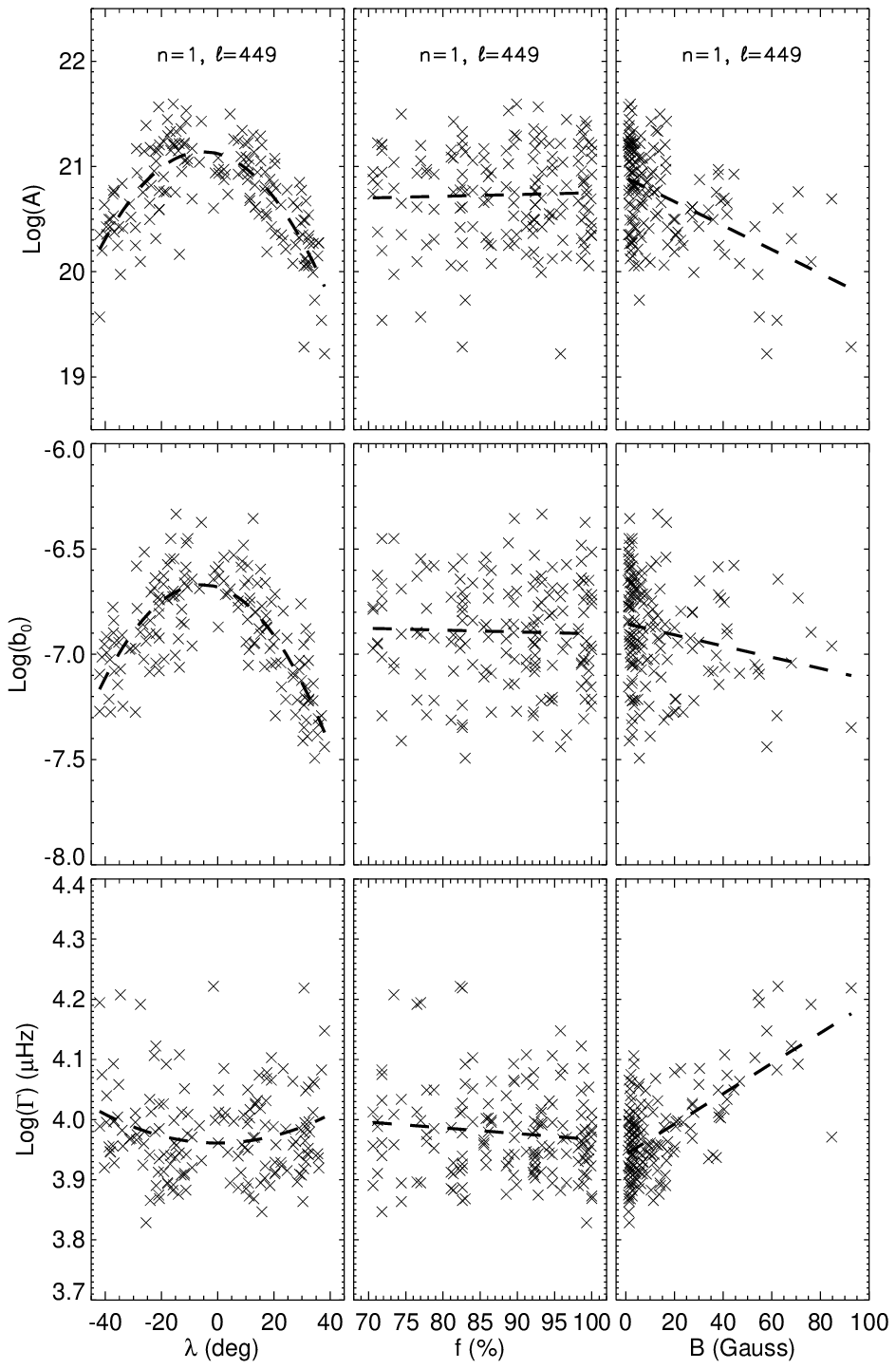}
    \caption{Mode parameters, amplitude $A$(top row), background power $b_0$ (middle row), and width $\Gamma$(bottom row) for two modes ($n=0,\ell=652$ (left three columns); $n=1,\ell=449$(right three columns)) for dormant ARs and QRs, as a function of longitude $\lambda$ of regions' centres (left), filling factor $f$ (middle), and  magnetic activity index $B$ (right). Dashed curves show the quadratic fit (in $\lambda$) and linear fit (in $f$ and $B$) through the mode parameters.}
    \label{fig:mod_var_mai_fill_dist}
\end{figure*}

\begin{figure*}[ht]
    \centering
        \includegraphics[width=1.0\textwidth,clip=,bb=54 29 465 344]{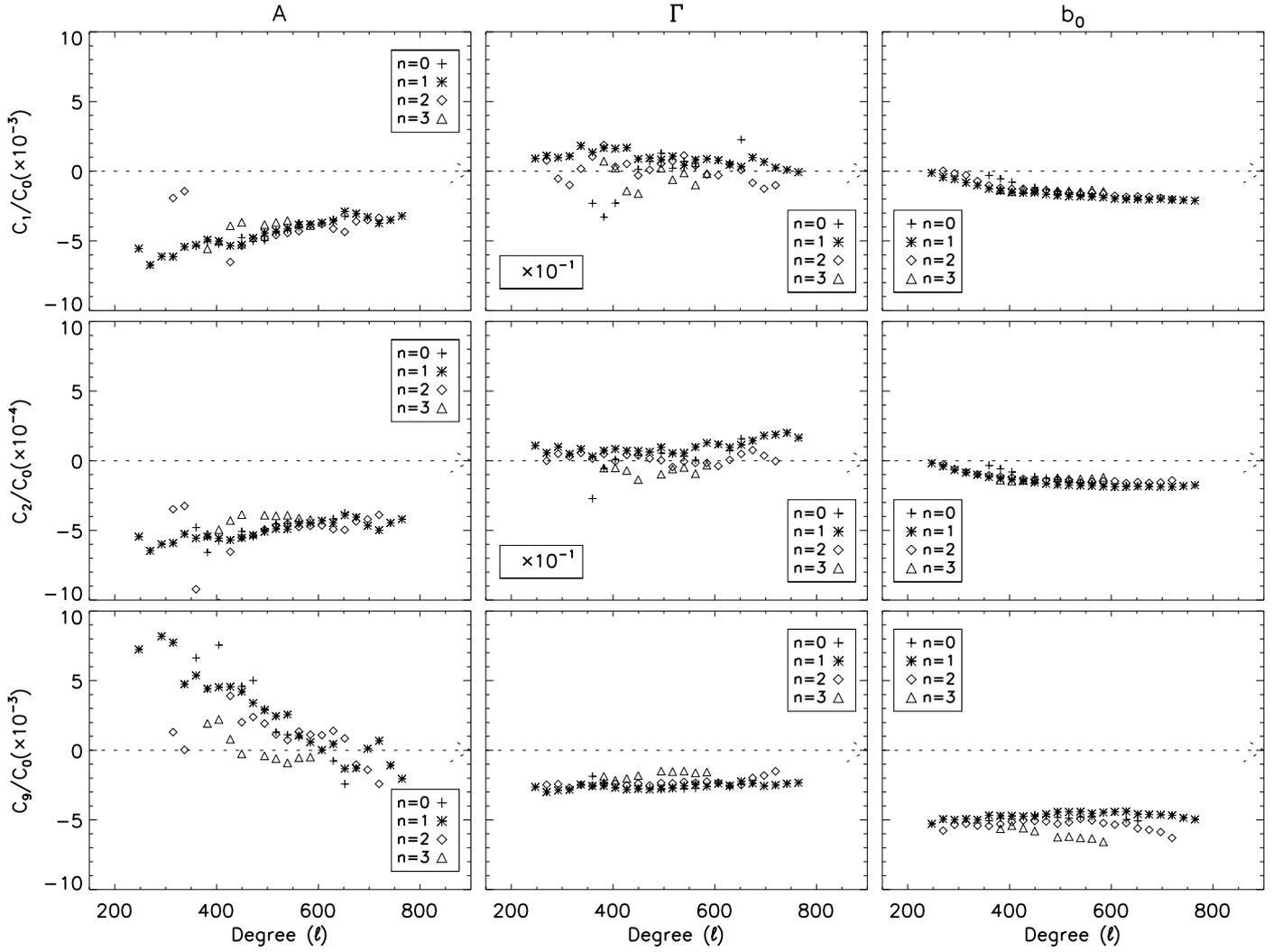}
    \caption{Multiple non-linear regression coefficients, $C_1$ (top row), $C_2$ (middle row) and $C_9$ (bottom row), normalized with intercept, $C_0$, for the mode parameters, $A$ (left column), $\Gamma$ (middle column) and $b_0$ (right column) as a function of harmonic degree, $\ell$. Corresponding correlation coefficients are shown in the Figure~\ref{fig:foreshort_correl_coef}.}
    \label{fig:foreshort_corr_coef}
\end{figure*}

\begin{figure*}[ht]
    \centering
        \includegraphics[width=1.0\textwidth,clip=,bb=54 29 465 344]{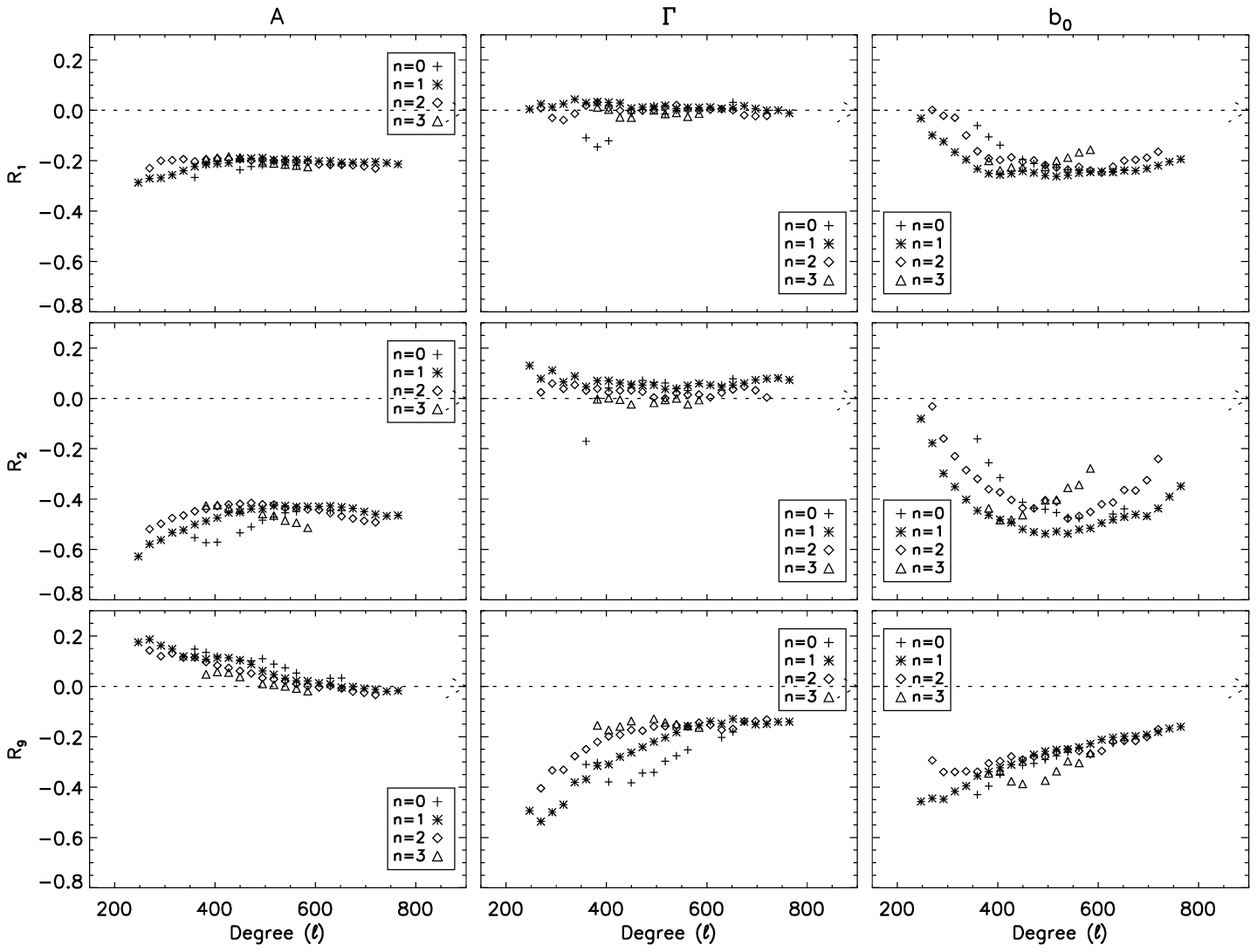}
    \caption{Correlation coefficients, $R_1$ (top row), $R_2$ (middle row) and $R_9$(bottom row) corresponding to first and second order terms in longitude $\lambda$ (see Equation~\ref{eq:foreshort}), respectively, for the mode parameters, $A$ (left column), $\Gamma$ (middle column) and $b_0$ (right column) as a function of harmonic degree, $\ell$.}
    \label{fig:foreshort_correl_coef}
\end{figure*}

\subsection{Mode Corrections}
\label{subsec:ModCorrect}

For a given multiplet ($n,\,\ell$), mode parameters change from region-to-region and with time. These changes are the combined effects of foreshortening, filling factor, magnetic activity, flare activity, and measurement uncertainties. In order to study the activity related changes in oscillation modes, we need to analyse and correct for the other effects.

Here filling factor is the filling of the observed Dopplergrams in the data cube used for p-mode computation by ring-diagrams. In our data samples, filling factor varies in the range 70 -- 100\%. However, most of the data cubes have filling factors > 80\%. Note that we are analysing the relative variations in mode parameters of ARs and corresponding QRs (e.g., Equation~\ref{eq:ModeEn-ratio}) which have same filling factors.

The Dopplergrams are significantly affected by foreshortening effects, i.e., as we go away from the disc centre, we measure only the cosine component of the vertical displacement. Another effect of foreshortening in mode parameters is caused by the fact that the spatial resolution in Dopplergrams decreases as we observe increasingly towards the limb. For example, a pixel which has a spatial resolution of $dx$ at the disk centre, now images a horizontal distance on the Sun of $dx/\rho$, where $\rho=\cos\Lambda\sin\Theta$. This reduces the spatial resolution measured on the Sun in the centre-to-limb direction, and hence leads to systematic errors.

Since we want to compare the mode parameters in different ARs, we analyse only the common modes among them. A mode is called common if all the data sets of ARs and QRs have same radial order ($n$) and degree ($\ell$). We have found total 98 common modes in our sample of ARs and QRs while total number of fitted modes ranges 125 -- 173. Figure~\ref{fig:mod_var_mai_fill_dist} shows examples of changes in common mode parameters of all the dormant ARs and QRs with longitude ($\lambda$), filling factor ($f$), and magnetic activity index ($B$).

One can infer from Figure~\ref{fig:mod_var_mai_fill_dist} that the mode amplitude ($A$) and background power ($b_0$) are more affected with distance from the disc centre than the mode width ($\Gamma$). The mode amplitude and background power decrease with increasing longitude while mode width shows an opposite relation. The variations in mode parameters with $\lambda$ are best modelled by quadratic function in $\lambda$ (see quadratic fit in Figure~\ref{fig:mod_var_mai_fill_dist}). The variations in background power with distance from the disc centre are stronger than the changes with the magnetic activity index. It is also to note that the measured longitude and latitude are affected by the position angle ($P$-angle) and $B_0$-angle of the Sun.

Changes in mode parameters with filling factors are shown in Figure~\ref{fig:mod_var_mai_fill_dist} (middle columns). It is seen that mode amplitude increases with increasing filling factor in the data cubes, while the mode width and background power show opposite trend. However, it is evident that the filling factor related variations in mode parameters are very small compared to the variations due to the foreshortening and magnetic activity indices.

We modelled the effects of foreshortening and filling factors on a common mode parameter ($\mathcal{P}_{n,\ell}$) of all the ARs and corresponding QRs by a multiple non-linear regression, as a function of longitude ($\lambda$) and latitude ($\theta$) of the centres of ring patches, position angle ($P$) and B-angle ($B_0$) of the Sun, and the filling factor ($f$) of data cubes,

\begin{eqnarray*}
\mathcal{P}_{n,\ell}(\lambda,\,\theta,\,\lambda_0,\,\theta_0,\,f)&=& C_0 + C_1\,\lambda + C_2\,\lambda^2 + C_3\,\theta + C_4\,\theta^2 \\
&&+ C_5\,P+ C_6\,P^2+C_7\,B_0+C_8\,B_0^2+C_9\,f
\label{eq:foreshort}
\end{eqnarray*}

\noindent where $C_0$, $C_1$, \ldots, $C_9$ are regression coefficients. We repeat this analysis for all the common pairs ($n,\ell$) of the ARs and QRs.

\begin{figure*}[ht]
    \centering
        \includegraphics[width=0.8\textwidth,clip=,bb=65 29 433 371]{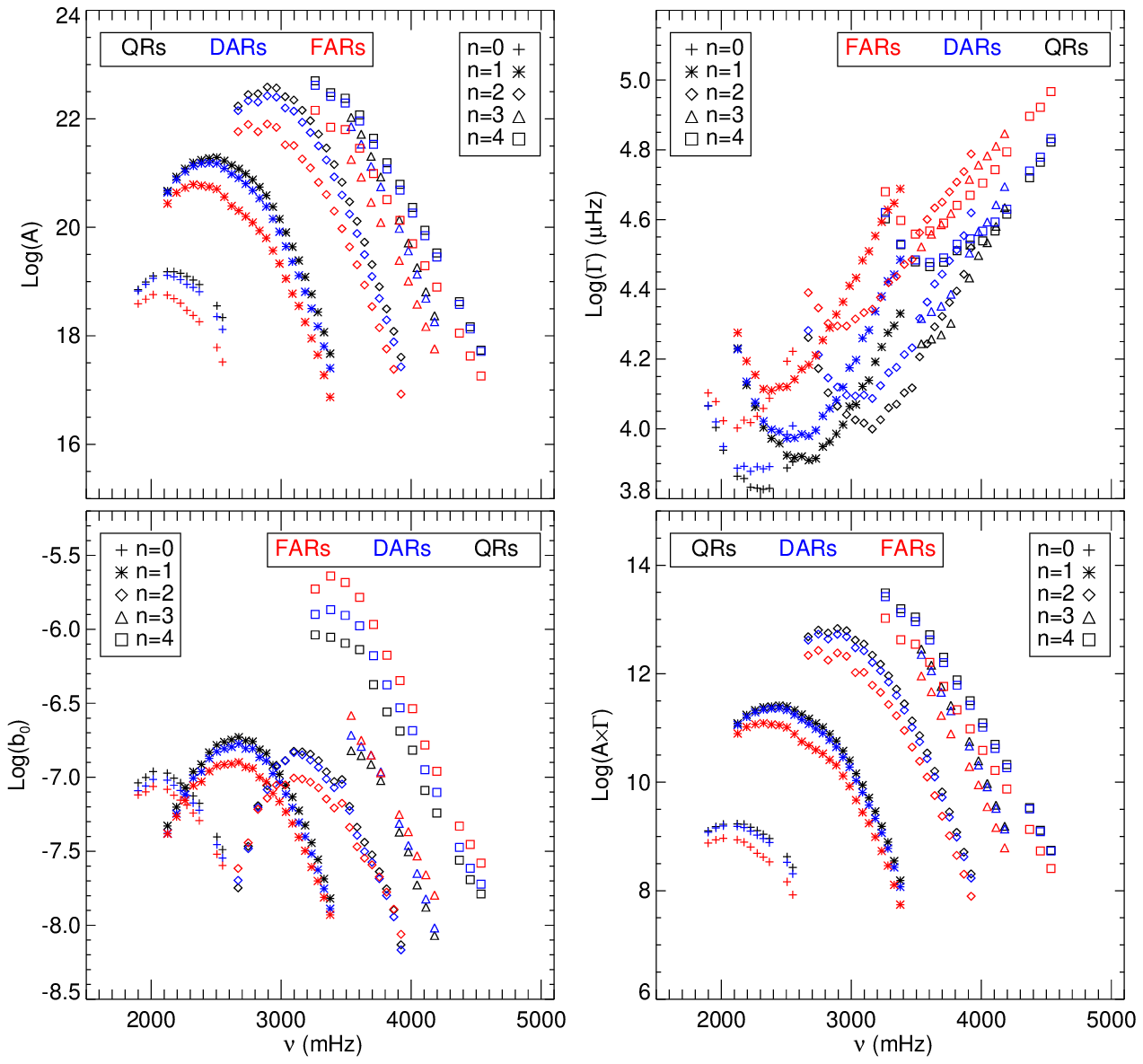}
    \caption{Average common mode parameters in QRs (black), dormant ARs (DARs; blue), and flaring ARs (FARs; red) for different radial orders as a function of mode frequency. Note that mode parameters are corrected for foreshortening and filling factor before averaging.}
    \label{fig:av_mode_par}
\end{figure*}

\subsubsection{Effects of the Foreshortening on the Mode Parameters}

Figure~\ref{fig:foreshort_corr_coef} shows the multiple non-linear regression coefficients ($C_1$, $C_2$ and $C_9$), normalized by the intercept $C_0$, for the common modes in all ARs and QRs, as a function of harmonic degree, $\ell$. The corresponding Pearson correlation coefficients are shown in Figure~\ref{fig:foreshort_correl_coef}. The  coefficients $C_2$ of the second order term for all the mode parameters are smaller than the coefficients $C_1$ of the first order term by one order of magnitude. But the correlation coefficients $R_2$ for all mode parameters are twice larger than the coefficients $R_1$ suggesting better association between the mode parameters and the quadratic term in $\lambda$. Both the coefficients $C_1$ and $C_2$ are negative for the mode parameters, $A$ and $b_0$, for all the harmonic degrees. For the mode width $\Gamma$, values of $C_1$ are positive for most of the modes with radial order $n=0,\,1$ and negative for $n=2,\,3$.  Also, magnitude of the coefficients for $A$ and $b_0$ are larger than that for $\Gamma$ for all harmonic degrees $\ell$. This shows that the mode amplitude and background power decrease rapidly while mode width increases slowly with increasing distance from the disc centre. Similar changes are also seen in the corresponding plots for correlation coefficients, $R_1$ and $R_2$ (Figure~\ref{fig:foreshort_correl_coef}).

Regression coefficients $C_1$ and $C_2$ for $A$ increase in magnitude with increasing harmonic degree ($\ell$), but mode width does show significant relation with $\ell$ in regression as well as in correlation coefficients. Figure~\ref{fig:foreshort_correl_coef} shows stronger anti-correlation between quadratic term in the mode parameter $A$ and in $\lambda$ than that of the linear term. There is a small positive correlation between the distance from the disc centre and mode width. But magnitudes of the correlation are very small,  $<0.10$. Regression coefficients $C_1$ and $C_2$ for background power decrease with increasing $\ell$; also evident from the significant anti-correlation between $b_0$ and $\lambda$. The correlation coefficients for background power changes with $\ell$, and radial order $n$. Thus we find that the background power is predominantly function of harmonic degree ($\ell$), decreasing with increasing $\ell$.

Similarly, the effects of latitude ($\theta$), p-angle ($P$) and B-angle ($B_0$) on mode parameters can also be illustrated. The foreshortening corrected mode parameters ($\mathcal{P}^c_{n,\ell}$) are obtained by subtracting the 1st and 2nd order terms (see Equation~\ref{eq:foreshort}) from the common mode parameters ($\mathcal{P}_{n,\ell}$). The foreshortening effects on p-mode amplitude and width have also been studied earlier \citep[e.g.,][]{Howe2004}. Our results support their reports.

\subsubsection{Effects of Filling Factor on the Mode Parameters}

The filling factor coefficient $C_9$ for the mode amplitude $A$ decreases with increasing $\ell$ (Figures~\ref{fig:foreshort_corr_coef}). It is positive for almost all modes at lower degree ($\ell\lesssim600$). In the intermediate range of degree, few modes have negative $C_9$ but with smaller magnitudes. The positive value of the coefficient $C_9$ shows that the mode amplitude increases with increasing filling factor while negative $C_9$ correspond to opposite relation. The increase of $A$ with $f$ is obvious from the increase in signal samples (or number of observed Dopplergrams) in the data cubes, but the reverse is not clear. This may be attributed to the distribution of observed Doppler frames in the data cubes. However, the magnitude of the correlation coefficients $C_9$ are very small for degree $\ell>450$.

The regression coefficients $C_9$ for the mode width and background power is negative at all degrees (Figure~\ref{fig:foreshort_corr_coef}). The magnitude of $C_9$ is smallest at lower $\ell$ and decreases with increasing $\ell$. Figure~\ref{fig:foreshort_correl_coef} also shows  anti-correlation between the mode width and filling factor. But the anti-correlation coefficients decrease with increasing harmonic degree $\ell$. The above values of $C_9$ and $R_9$ show that the mode width and background power decrease with increasing filling factor.

\begin{figure*}[ht]
    \centering
        \includegraphics[width=0.49\textwidth,clip=,bb=17 7 346 465]{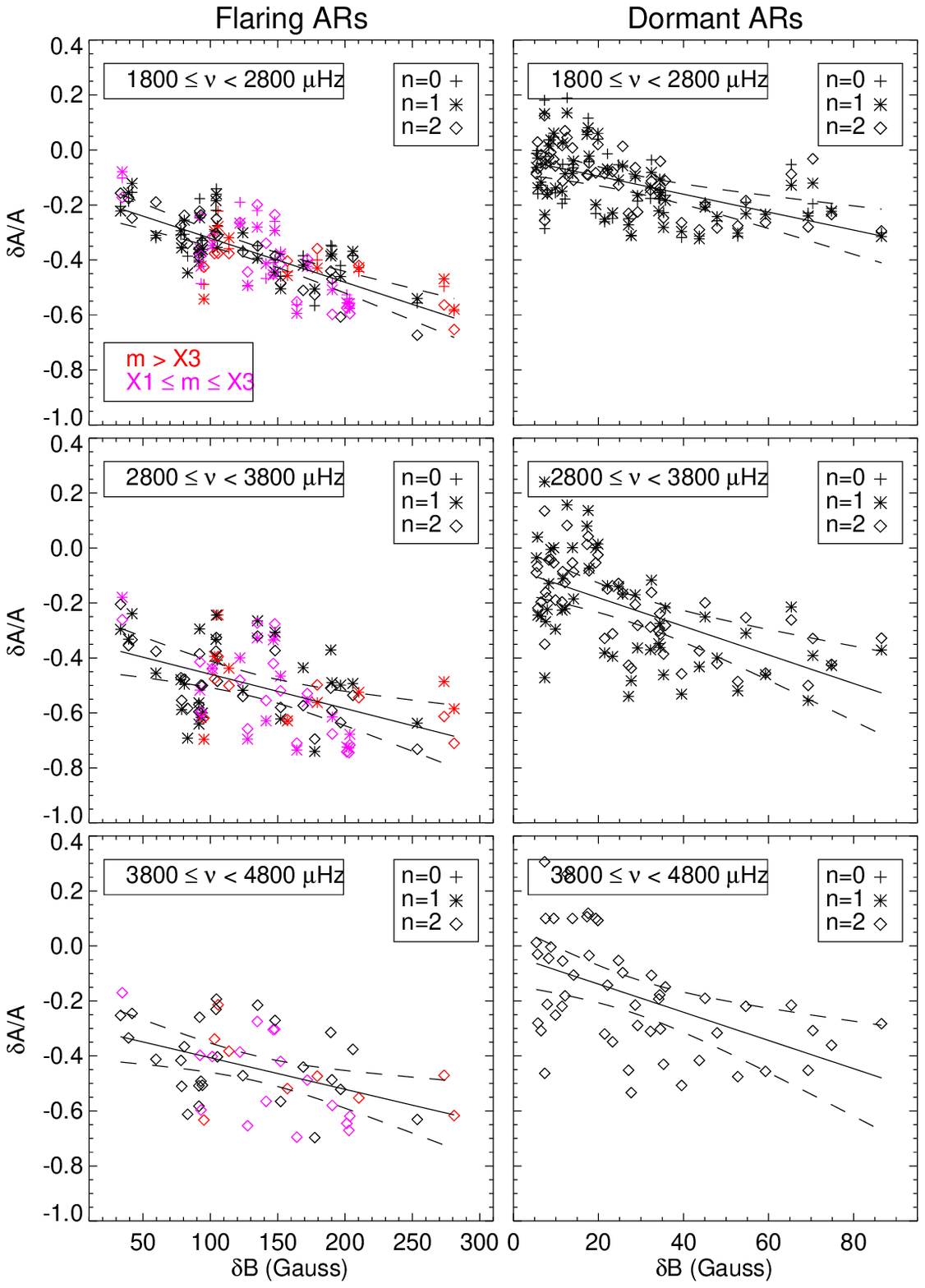}
        \includegraphics[width=0.49\textwidth,clip=,bb=17 7 346 465]{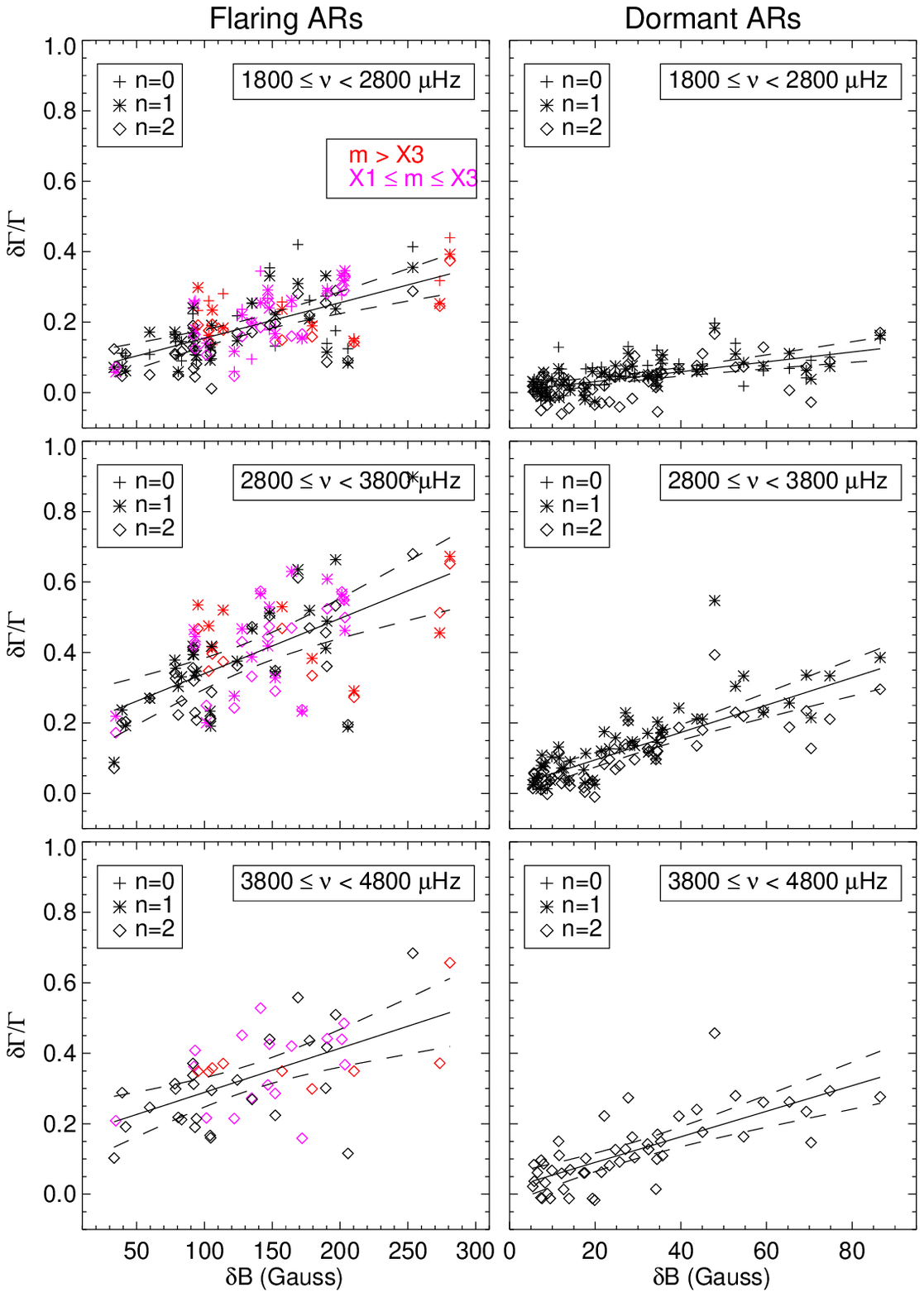}
    \caption{Frequency averages of the fractional differences in mode amplitude (left two columns) and mode width (right two columns) for flaring (left) and dormant (right) ARs as a function of magnetic activity index difference ($\delta B=B_{\rm AR}-B_{\rm QR}$) of AR and corresponding QR. Red and magenta colours symbols correspond to those ARs which have flare(s) of magnitude (say, $m$) > X3 and X1$\leq m\leq X3$, respectively. Solid lines show the linear regression fit while dashed curves show 90\% confidence level of the linear fit. Note that the mode parameters are corrected for the foreshortening and filling factors.}
    \label{fig:freq_av_ap_wd_mai}
\end{figure*}

\begin{figure*}[ht]
    \centering
        \includegraphics[width=0.49\textwidth,clip=,bb=17 7 346 465]{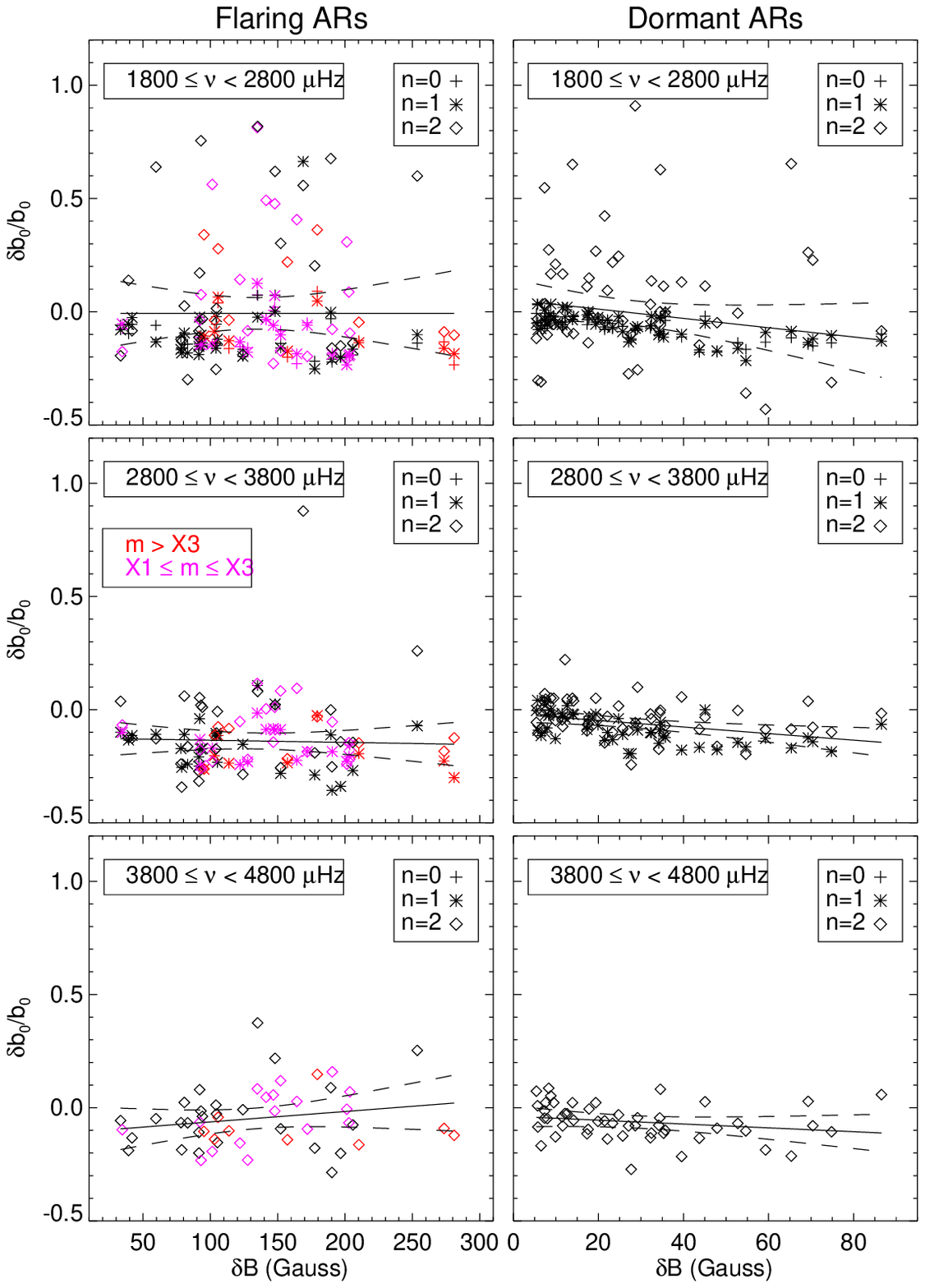}
        \includegraphics[width=0.49\textwidth,clip=,bb=17 7 346 465]{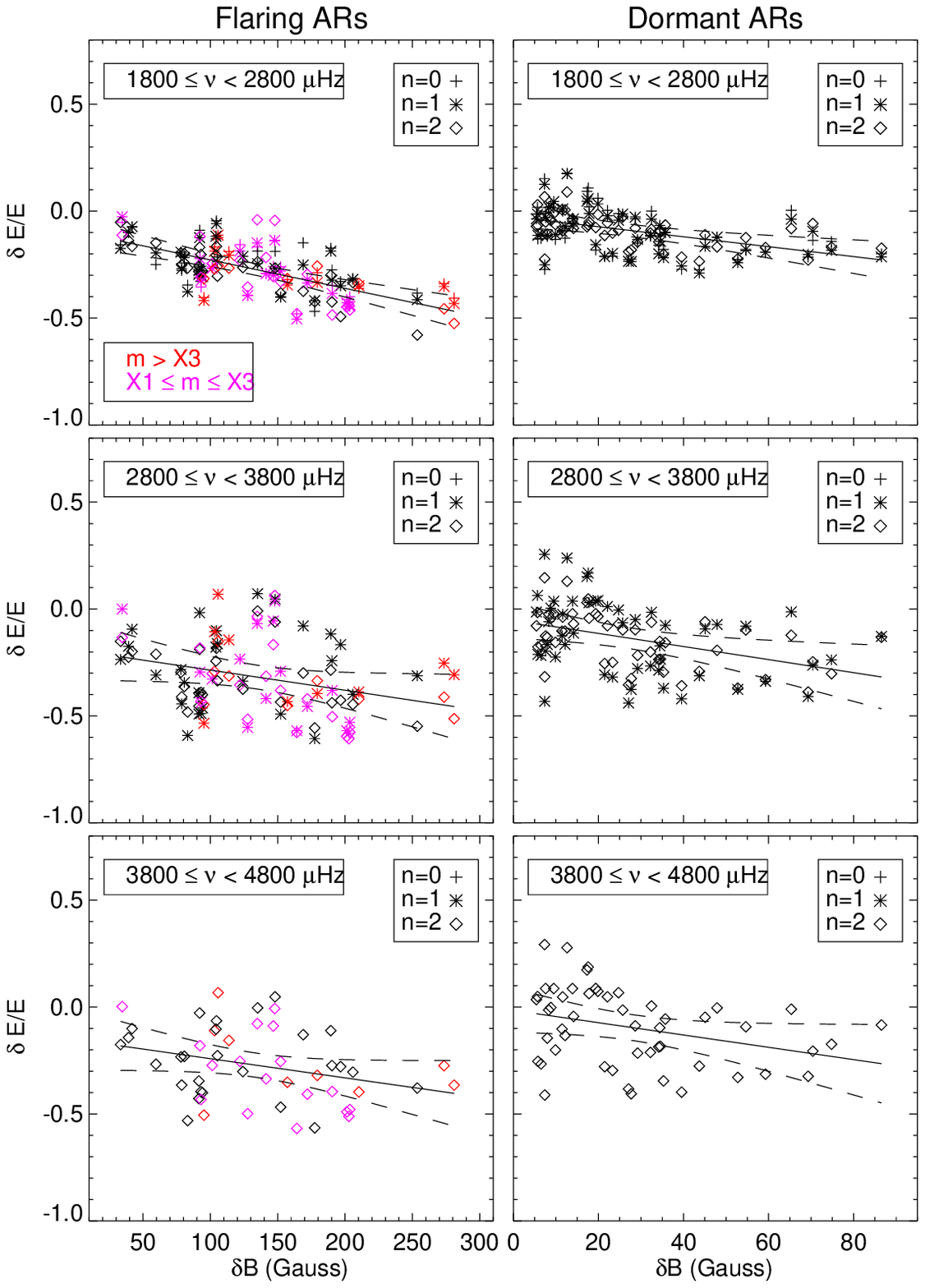}
    \caption{Similar to Figure~\ref{fig:freq_av_ap_wd_mai} but for the background power (left two columns) and mode energy (right two columns).}
    \label{fig:freq_av_b0_area_mai}
\end{figure*}

\section{Results and Discussions}
\label{sec:ResultsDiscs}

\subsection{Variation in p-Mode Parameters with Magnetic Activity Index}
\label{subsec:ModPar-MAI}

In order to analyse the variation in mode parameters with magnetic activity index ($B$) of flaring and dormant ARs, we apply the mode corrections for  foreshortening and filling factor to all the mode parameters of all the ARs and QRs.

Figure~\ref{fig:av_mode_par} shows the averaged mode parameters in QRs, dormant ARs and flaring ARs as a function of frequency. Evidently, the average mode amplitudes in QRs are larger than those in dormant and  flaring ARs for all radial orders. Flaring ARs, possessing largest values of B, show smallest average mode amplitude in all radial orders. This is obviously due to the larger mode power suppression in ARs, having stronger magnetic activity index ($B$), than in the QRs. Average background power for $n\le2$ in ARs and QRs varies with $B$ similar to the average mode amplitude. For $n\geq3$, background power is largest in flaring ARs and smallest in QRs. Average mode widths are largest in flaring ARs and smallest in QRs for all radial orders, showing shorter lifetimes of modes in flaring ARs than in dormant ARs and QRs. Mode area ($A\times\Gamma$) of ARs and QRs shows similar trend as the mode amplitude.

Figures~\ref{fig:freq_av_ap_wd_mai} and~\ref{fig:freq_av_b0_area_mai} show frequency averages of the fractional difference in mode parameters as a function of magnetic activity index difference ($\delta B=B_{\rm AR}-B_{\rm QR}$), between ARs and corresponding QRs, for flaring and dormant ARs in three frequency bands, as indicated. The coefficients of linear regression and Pearson correlation between the frequency averaged fractional mode differences in the three frequency bands and $\delta B$ are given in Table~\ref{tab:mod-ratio-mai}. The averaging of mode parameters in these frequency bands, however, would somewhat hide the frequency dependent properties of modes.

\begin{figure*}[ht]
    \centering
        \includegraphics[width=0.49\textwidth,clip=,bb=14 6 241 501]{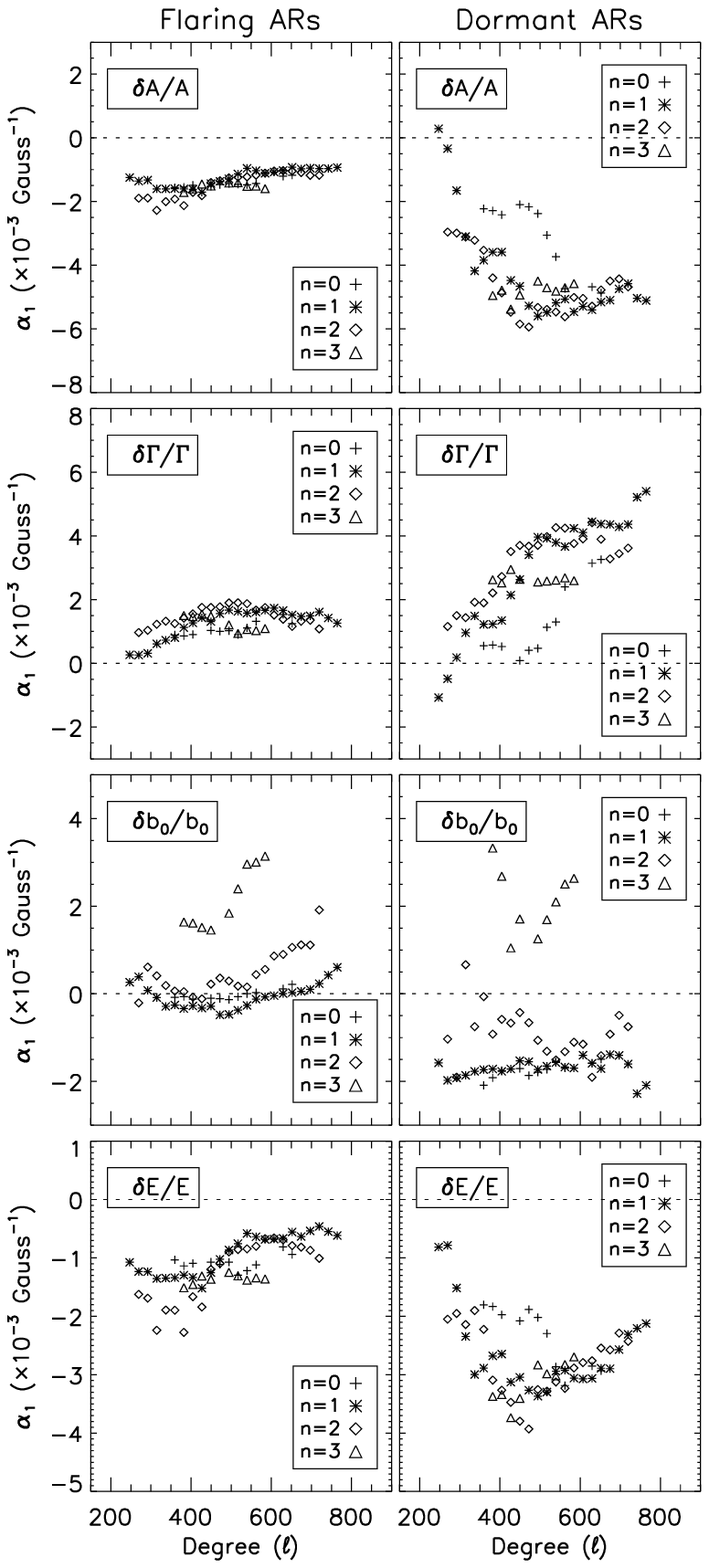}
        \includegraphics[width=0.49\textwidth,clip=,bb=14 6 241 501]{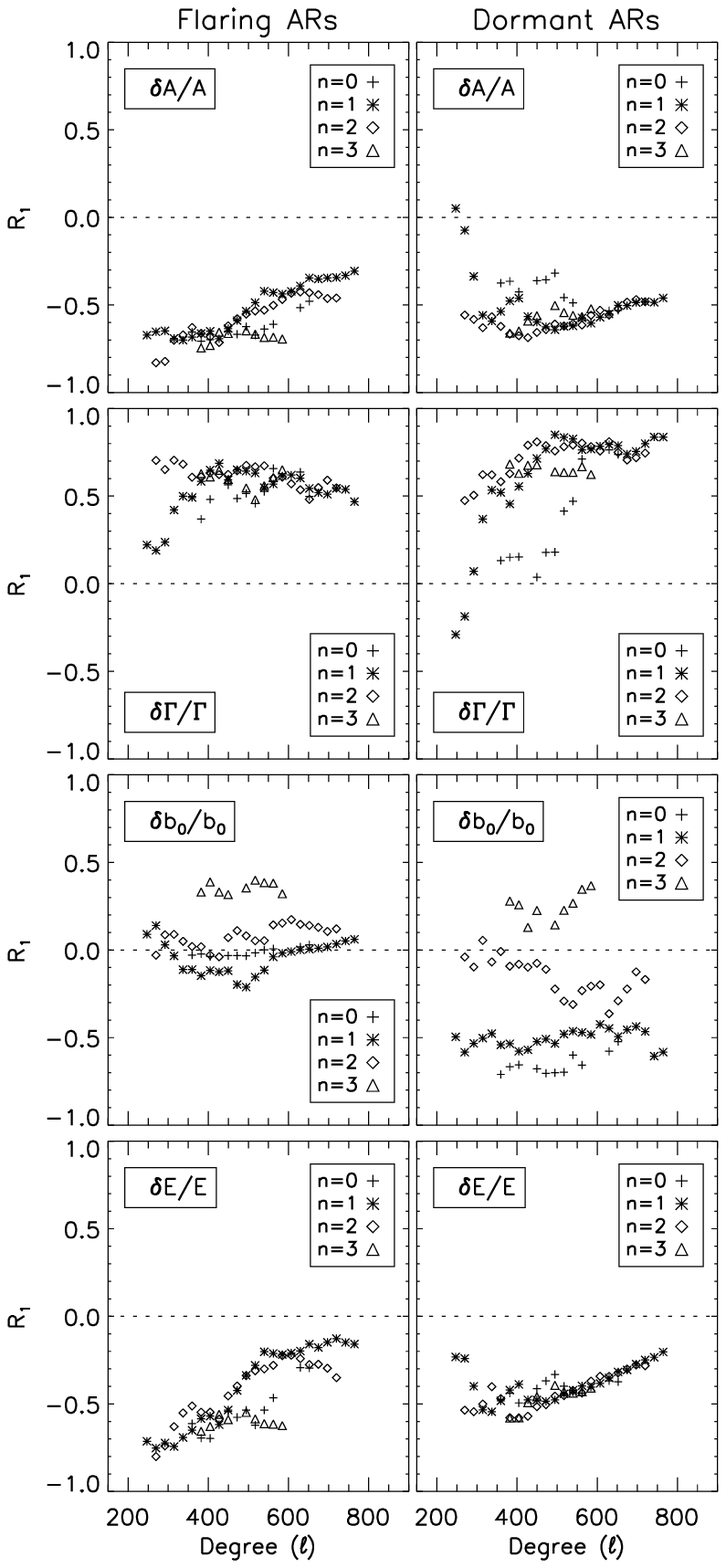}
    \caption{Coefficients of linear regression in $\delta B$ (left two columns) and Pearson correlation (right two columns), for different mode parameters of flaring and dormant ARs, as function of harmonic degree.}
    \label{fig:freq_mai_cf}
\end{figure*}

\begin{figure*}[ht]
    \centering
        \includegraphics[width=0.49\textwidth,clip=,bb=14 6 241 501]{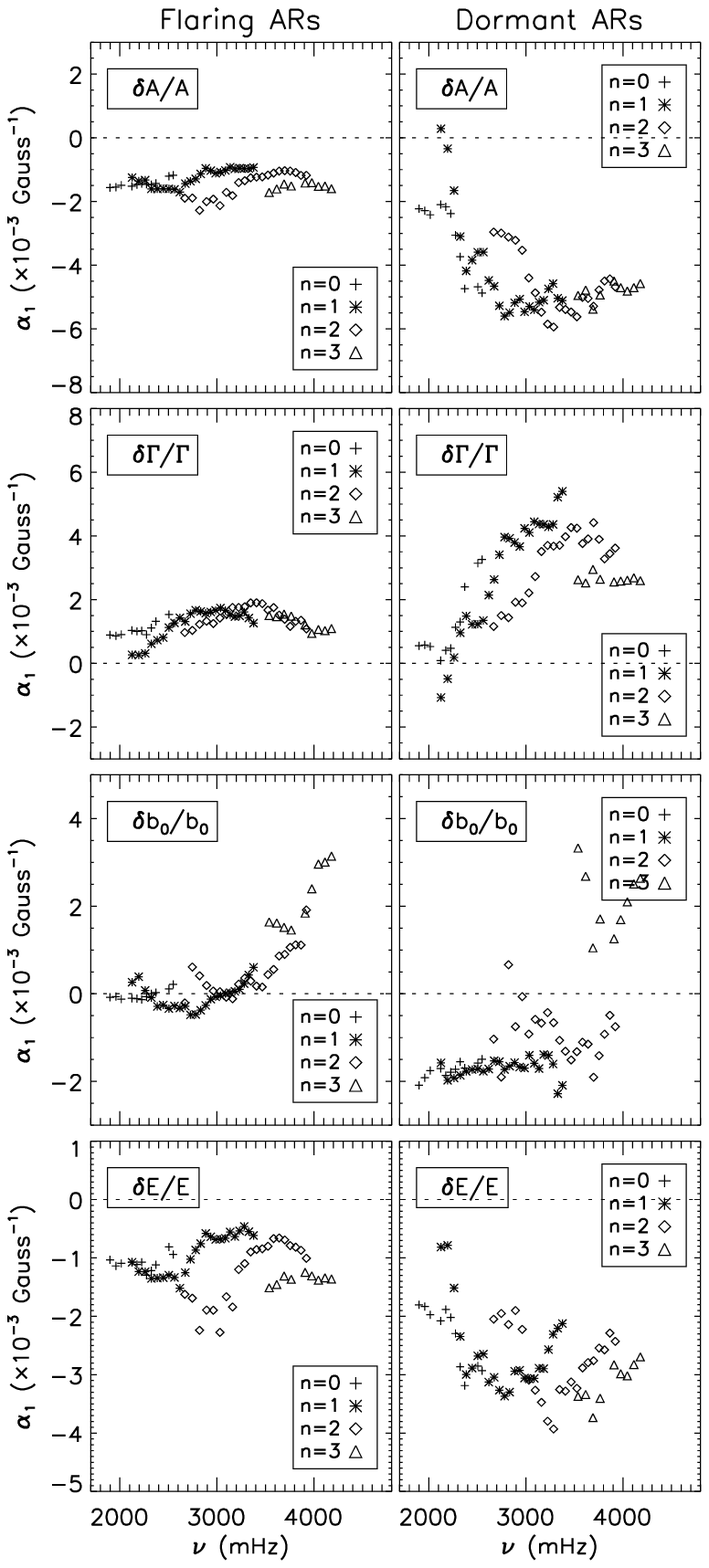}
        \includegraphics[width=0.49\textwidth,clip=,bb=14 6 241 501]{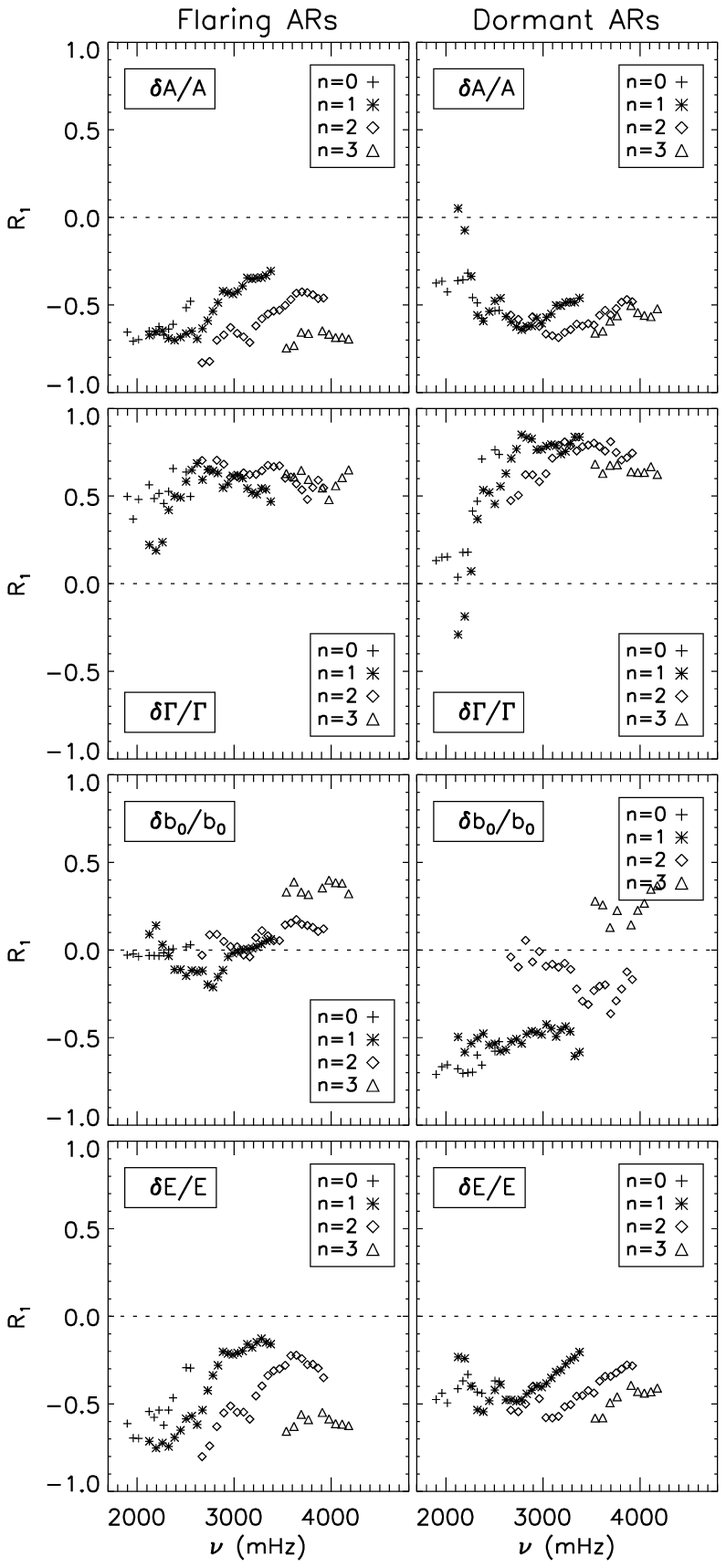}
    \caption{Similar to Figure~\ref{fig:freq_mai_cf} but as a function of frequency.}
    \label{fig:freq_mai_cf_nu}
\end{figure*}

Previous statistical studies of several ARs showed that mode amplitude and width are linearly related \citep[e.g.,][]{Howe2004, Rabello-Soares2008}. Therefore, to find the harmonic degree and frequency dependent relation between mode parameters and $\delta B$, we fit the fractional mode differences (say, $\mathcal{R}$) and $\delta B$ with linear regression, $\mathcal{R}_{n,\ell}(\delta B)=\alpha_0+\alpha_1\,(\delta B)$. Slope ($\alpha_1$) of the linear regression will then correspond to the mode parameter variation per Gauss, hereafter the ``parameter variation rate''. The coefficients $\alpha_1$ of the linear regression for different mode parameters are shown in Figures~\ref{fig:freq_mai_cf} and~\ref{fig:freq_mai_cf_nu} respectively as a function of harmonic degree ($\ell$) and frequency ($\nu$). The constant term $\alpha_0$ has no intrinsic meaning and is not illustrated. In the following, we discuss various mode parameters.

\begin{table*}[ht]
\centering
\caption{Linear regression ($y=a+b\,\delta B$) and Pearson correlation coefficients ($r$) obtained from the fractional mode differences ($y$) in three frequency bands and magnetic activity indices ($\delta B$). Note that the mode parameters are corrected for the foreshortening and the filling factors.}
\begin{tabular}{|l|l|r|r|r|r|r|r|}
\hline
                                  &            &   \multicolumn{3}{c}{Flaring ARs} &  \multicolumn{3}{|c|}{Dormant ARs}\\
\cline{3-8}
                       Mode      &      $\nu$ -range &  $a$ &   $b$ $\pm$          $\sigma$ &   $r$ &   $a$ &   $b$ $\pm$          $\sigma$ &   $r$\\
\cline{2-8}
                   Parameters               & ($\mu Hz$)        &      &            $(\times10^{-3} {\rm Gauss}^{-1})$ &       &       &            $(\times10^{-3} {\rm Gauss}^{-1})$ &      \\
\hline
\hline
                                  &       1800 - 2799 &-0.16 & -1.60 $\pm$              0.18 & -0.78 & -0.03 & -3.27 $\pm$              0.63 & -0.60\\
                     $\delta A/A$ &       2800 - 3799 &-0.33 & -1.24 $\pm$              0.29 & -0.52 & -0.08 & -5.20 $\pm$              0.97 & -0.61\\
                                  &       3800 - 4800 &-0.29 & -1.15 $\pm$              0.31 & -0.46 & -0.04 & -5.13 $\pm$              1.23 & -0.51\\
\hline
                                  &       1800 - 2799 &-0.01 &  0.00 $\pm$              0.48 &  0.00 &  0.05 & -2.05 $\pm$              1.05 & -0.27\\
                 $\delta b_0/b_0$ &       2800 - 3799 &-0.13 & -0.10 $\pm$              0.24 & -0.06 & -0.02 & -1.45 $\pm$              0.40 & -0.46\\
                                  &       3800 - 4800 &-0.11 &  0.46 $\pm$              0.31 &  0.20 & -0.04 & -0.85 $\pm$              0.53 & -0.22\\
\hline
                                  &       1800 - 2799 & 0.05 &  1.01 $\pm$              0.14 &  0.71 &  0.00 &  1.40 $\pm$              0.21 &  0.69\\
            $\delta\Gamma/\Gamma$ &       2800 - 3799 & 0.18 &  1.57 $\pm$              0.26 &  0.65 &  0.02 &  3.89 $\pm$              0.37 &  0.83\\
                                  &       3800 - 4800 & 0.16 &  1.25 $\pm$              0.24 &  0.58 &  0.02 &  3.63 $\pm$              0.47 &  0.74\\
\hline
                                  &       1800 - 2799 &-0.10 & -1.33 $\pm$              0.19 & -0.71 & -0.03 & -2.33 $\pm$              0.56 & -0.51\\
                     $\delta E/E$ &       2800 - 3799 &-0.19 & -0.94 $\pm$              0.38 & -0.33 & -0.05 & -3.06 $\pm$              0.95 & -0.42\\
                                  &       3800 - 4800 &-0.15 & -0.90 $\pm$              0.39 & -0.31 & -0.02 & -2.88 $\pm$              1.18 & -0.33\\
\hline
\end{tabular}

\label{tab:mod-ratio-mai}
\end{table*}

\subsubsection{Mode Amplitude}

Figures~\ref{fig:freq_av_ap_wd_mai} (left two columns) shows frequency averaged fractional differences in mode amplitudes ($\delta A/A$) in three frequency bands. For both flaring and dormant ARs, $\delta A/A$ are below zero, i.e., negative. However, they are more negative for flaring ARs than for dormant ARs. The amplitude difference decreases with increasing $\delta B$ which is further confirmed by the significant correlation coefficients ($r$) between $\delta A/A$ and $\delta B$ (see Table~\ref{tab:mod-ratio-mai}). Magnitude of correlation coefficients are larger in the lower frequency band than in higher frequency bands for flaring ARs. In the flaring ARs, power suppression per unit Gauss is larger than in dormant ARs as evident from the linear regression slope $b$ (see Table~\ref{tab:mod-ratio-mai}). Furthermore, in the five minute band of dormant ARs, we infer a larger suppression than in the lower and higher frequency bands.

The amplitude variations for both flaring and dormant ARs are well correlated with $\delta B$ as evident from the magnitude of large Pearson correlation coefficients (see Figures~\ref{fig:freq_mai_cf} and~\ref{fig:freq_mai_cf_nu}). This confirms linear relation between amplitude variation and $B$ \citep{Rabello-Soares2008}.

Figure~\ref{fig:freq_mai_cf} and~\ref{fig:freq_mai_cf_nu} show changes in linear regression coefficient $\alpha_1$ for mode amplitude (i.e., rate of change of $\delta A/A$ per unit Gauss) as a function of harmonic degree ($\ell$) and frequency ($\nu$), respectively. For the flaring ARs, regression coefficients $\alpha_1$ increase slowly with harmonic degree $\ell$. Also, correlation coefficients decrease in magnitude with increasing $\ell$. Furthermore, correlation coefficients for both flaring and dormant ARs change with harmonic degree. For the dormant ARs, magnitude of correlation coefficients for different radial orders first decreases with harmonic degree ($\ell$), becomes smallest at $\ell$ in the range $\sim$450-600, and then increases again. This is consistent with \citet{Bogdan1993} who reported maximum absorption around wavenumber $k\approx0.8\,{\rm Mm}^{-1}$ or $\ell\approx550$. This value of $\ell$ approximately agrees with our values where we find maximum suppression of power for $n=1,\,2$ and 3 modes. For the flaring ARs,  correlation coefficients monotonically increase with degree, except for a few modes at smaller degree, for all the radial orders. The significant correlation coefficients further support the relationships between $\delta A/A$ and $\delta B$ of the flaring and dormant ARs.

Figure~\ref{fig:freq_mai_cf_nu} (first row from top) shows that amplitude change in dormant ARs with frequency is not monotonic. Magnitude of the regression coefficient $\alpha_1$ of dormant ARs decreases with frequency and becomes smallest in the five minute band then decreases. For flaring ARs, $\alpha_1$ does not show a significant frequency dependence.

The amplitude decrease rate  (i.e., $\alpha_1$) for dormant ARs increases with frequencies and peaks around 3.0\,mHz and 3.5\,mHz for the radial orders, $n=1$ and $n=2$, respectively. This supports the previous reports \citep{Rajaguru2001, Howe2004, Rabello-Soares2008}. For global modes, \citet{Komm2000} found $\sim 7$\% decrease in mode amplitude with the solar cycle. They also found that the maximum variations of 29\% occur in the frequency range of 2.7--3.3\,mHz. More interestingly, we find that the amplitude variation rate for dormant ARs increases at higher frequencies. The observed increase of the mode absorption as a function of frequency and decrease at higher frequencies have been theoretically reported earlier \citep{Jain2009a}. Negative values of $\alpha_1$ may be attributed with power absorption \citep{Braun1987}. But it is not clear if the mode power absorption can be compared with suppression of power in the sunspot region, because absorption was calculated for travelling waves while the modes we are analysing are standing waves.

Smaller values of the coefficients $\alpha_1$ for the flaring ARs than the dormant ARs may be the net result of both mode power absorption by sunspots and amplification by flares. However, the absorption effect clearly dominates as evident from the negative fractional difference in mode amplitude (Figure~\ref{fig:freq_av_ap_wd_mai}).

\subsubsection{Mode Width}

Mode width of a peak profile in the power spectrum is related to the imaginary part of frequency and hence is an indication of mode damping. It is inversely proportional to the lifetime of the mode. Frequency averaged fractional difference in mode width ($\delta\Gamma/\Gamma$) for flaring and dormant ARs as a function of magnetic activity index difference ($\delta B$) are shown in Figure~\ref{fig:freq_av_ap_wd_mai} (right two columns). $\delta\Gamma/\Gamma$ of both the flaring and dormant ARs are positive and increase with increasing $\delta B$ employing the decrease in lifetime of modes with $B$. In the five minute band of both the flaring and dormant ARs, $\delta\Gamma/\Gamma$ increases faster than in lower and higher frequency bands. This relation is further confirmed from the large correlation coefficients (see Table~\ref{tab:mod-ratio-mai}). But the correlations are better in the five minute band of the dormant ARs and in the lower frequency band of flaring ARs. The larger values of the intercept $a>0$ for dormant ARs than flaring ARs show that mode width is large in ARs than in QRs. Thus it appears that most modes live longer in QRs than they do in ARs, employing the larger damping in ARs. For dormant ARs, data points are more close to linear regression line while for flaring ARs there is large spread at larger $\delta B$ causing smaller correlations. This may be caused by activities in the flaring ARs. Furthermore, the larger slope in the five minute band of flaring ARs than in dormant ARs shows that mode width in flaring ARs decreases slowly with $\delta B$ than in dormant ARs.

The coefficients ($\alpha_1$) of linear regression between the fractional difference in mode width and $\delta B$ are shown in Figures~\ref{fig:freq_mai_cf} and~\ref{fig:freq_mai_cf_nu} (second row from the top) as a function of harmonic degree ($\ell$) and frequency ($\nu$), respectively. The coefficients $\alpha_1$ for both dormant and flaring ARs are positive and increase with $\ell$. Its values are smaller for flaring ARs than the dormant ARs for all $\ell$ except for a few modes with $\ell<350$. The maximum value of the width variation for flaring and dormant ARs occur in the range $\sim 1 - 2\,{\rm Gauss}^{-1}$ and $\sim 2 - 6\,{\rm Gauss}^{-1}$, respectively.

\citet{Chen1996}, using the absorption of p-mode waves in sunspot as a tool to determine the lifetime of p-modes in the $\ell$ range (200-700), reported that p-mode lifetimes decrease with $\ell$ and frequency ($\nu$). The width variation rate of dormant ARs initially increases with frequency for all radial orders, peaks at different frequencies in the five minute frequency band, and then decreases (except $n=0$). For radial order, $n=0$, the maximum $\sim0.0032\,{\rm Gauss}^{-1}$ occurs around $\nu\approx2.6$\,mHz. But it is to note that we did not cover the modes $\nu>2.6$ for $n=0$.  \citet{Rajaguru2001}, \citet{Howe2004}, and \citet{Rabello-Soares2008} have reported the maximum around 0.005, 0.0045 and 0.003 Gauss$^{-1}$, respectively.

For the global modes, \citet{Komm2000} reported solar cyclic mode width variations of about 3\% and the maximum changes of 47\% in the frequency range 2.7--3.5\,mHz. But they did not find $\ell$ dependence of solar cycle changes in mode width presumably because of smaller variations in the average global magnetic fields.

\subsubsection{Background Power}

The background power in solar oscillation spectra is not just ``noise'', but contains physical information which might be important for better understanding of the dynamics of the solar atmosphere. It has a large component of so-called solar noise, which is the background produced by convective cells. Furthermore, the detection of oscillation modes essentially depends upon the signal-to-noise ratio, therefore, it is important to compare the background noise in the power spectra between the ARs and corresponding QRs.

Figure~\ref{fig:freq_av_b0_area_mai} (left two columns) shows that the relative variation in frequency averaged background power of dormant ARs decreases slowly with increasing $\delta B$ while it is almost constant for flaring ARs. For most of the ARs, background power is less than their corresponding QRs which confirms earlier reports \citep{Rajaguru2001}. The reduction may be expected because magnetic field is known to suppress convection, which is the main source of background in power spectra. But the fractional difference in background power at lower frequency band is larger than zero for a  few modes with larger radial order, e.g., $n=2$ (see Figure~\ref{fig:freq_av_b0_area_mai}). The correlation between $\delta b_0/b_0$ and $\delta B$ is poor or no-correlation in flaring ARs while there is significant anti-correlation for dormant ARs (see Table~\ref{tab:mod-ratio-mai}). The poor correlation coefficient for flaring ARs may be caused by mode suppression by sunspots and amplification by flares.

\citet{Komm2000} found no significant variation in background power with solar cycle in the global modes. The expected variation in background power with solar cycle in their analysis may be due to small increase in the average magnetic field with the solar cycle. Also the global mode analysis is restricted to the $\ell<200$, where variation is small even in our results.

The coefficients of linear regression ($\alpha_1$) between the fractional difference in background power and $\delta B$ are shown in Figures~\ref{fig:freq_mai_cf} and~\ref{fig:freq_mai_cf_nu} (third row from the top) as a function of $\ell$ and $\nu$, respectively. The coefficient $\alpha_1$ is negative (positive) for most flaring (dormant) ARs with $n<2$. For p-modes of flaring ARs, it first decreases with frequency and becomes smallest at different frequencies, in the five minute band, for different radial orders, and then increases. Negative value of coefficients $\alpha_1$ further shows the opposite relation between background power and magnetic activity index, B. The positive value of $\alpha_1$ for flaring ARs shows power enhancement which may be caused by flare induced excitation. However, the correlation coefficients for most of the modes are very small for flaring ARs.

\begin{figure*}[ht]
    \centering
        \includegraphics[width=1.0\textwidth,clip=,bb=32 10 462 240]{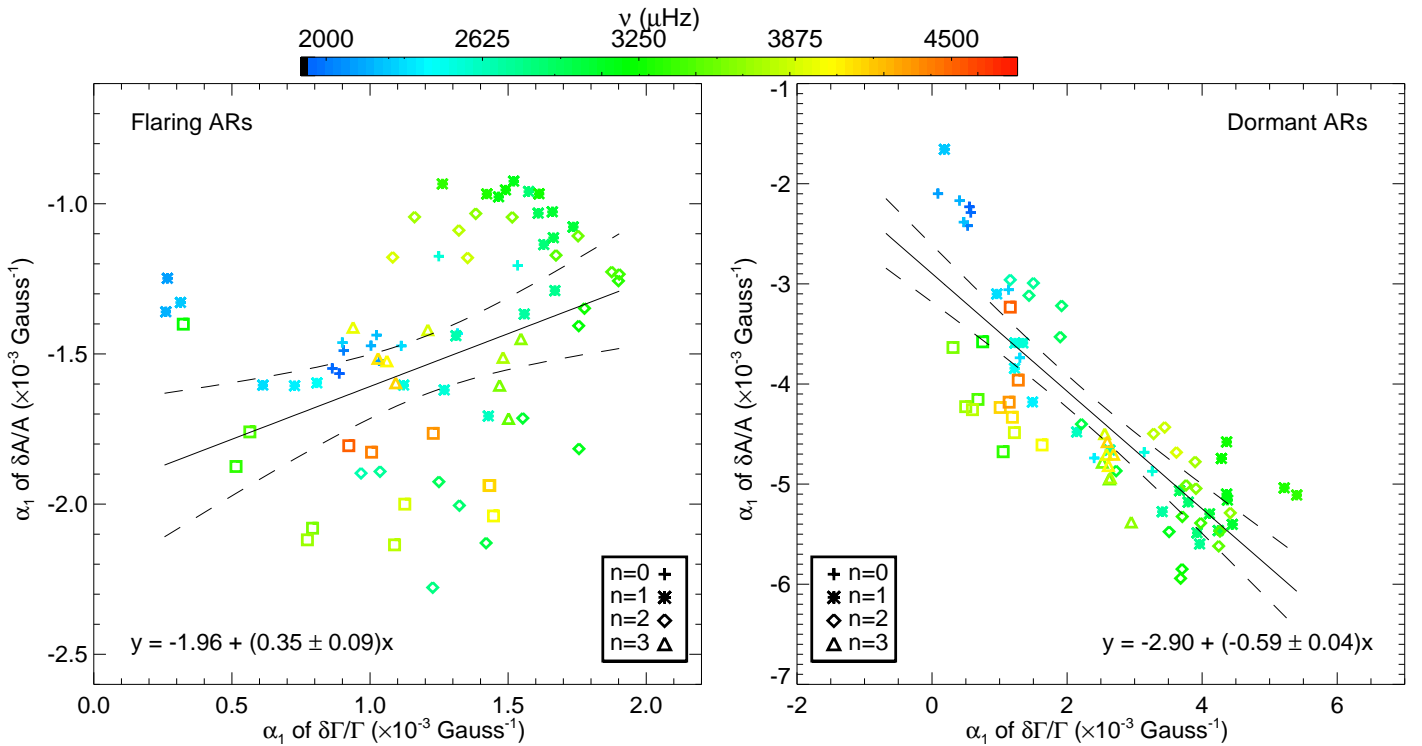}
    \caption{Variations in the rate of fractional differences in mode amplitude and mode width of flaring (left) and dormant (right) ARs. Solid lines show the linear regression fit with the fitted coefficients as given in the respective equations.}
    \label{fig:ap_vs_wd_mai}
\end{figure*}

\subsubsection{Mode Energy}

If the total power in the mode is the area under the peak in the power spectrum, it is not instantly clear whether the increase in line widths in ARs compensates for the decrease in peak height. The area under a peak is a measure of excitation for the corresponding mode. To examine these changes, we plotted the fractional difference of the product of amplitude and width as a proxy for the area under the peak or mode energy (see Equation~\ref{eq:ModeEn-ratio}) in three frequency bands. This is shown in Figure~\ref{fig:freq_av_b0_area_mai} (right two columns).

Figure~\ref{fig:freq_av_b0_area_mai} shows that the mode energy for both flaring and dormant ARs are smaller than in the corresponding QRs and decreases with increasing $\delta B$. But the decrease rate in dormant ARs is faster than in flaring ARs as evident from the linear regression slope (see Table~\ref{tab:mod-ratio-mai}). In the dormant ARs the energy difference decreases faster in the five minute band than in the lower and higher frequency bands. The intercept ($a$), and also, the anti-correlation coefficient decrease from lower to higher frequencies. For the global modes, \citet{Komm2000} reported a reduction of up to 60\% in mode area. They found that maximum reduction of 36\% in mode area occurs in the frequency range of 2.7--3.3\,mHz.

The linear regression fit shows that the slopes of the flaring ARs are very small for five minute band. This implies that mode energy slowly decreases with $B$ of flaring ARs. This is perplexing as several studies have reported strong p-mode power absorption by ARs' magnetic fields  \citep{Braun1987, Braun1990,Rajaguru2001,Mathew2008}. Therefore, mode energy is expected to decrease faster with increasing $B$ rather than with the rate as inferred by the small slope for energy difference for the flaring ARs in Figure \ref{fig:freq_av_b0_area_mai}.

Figure~\ref{fig:freq_mai_cf} and~\ref{fig:freq_mai_cf_nu} (bottom panel) show the coefficients of linear regression $\alpha_1$ between fractional mode energy difference $\delta E/E$ and $\delta B$ as a function of harmonic degree and frequency, respectively. The energy variation rate (i.e., $\alpha_1$) for dormant ARs initially decreases with increasing degree $\ell$, and becomes smallest around $\ell\approx500$ for all the radial orders then increases. Previous studies have shown decrease in mode energy with harmonic degree at fixed frequency \citep[e.g.][]{Rhodes1991}. \citet{Woodard2001} have also studied the energy rate in the intermediate mode. They infer that the time-averaged energy per mode, which is theoretically related to the modal surface velocity power, decreases steeply with $\ell$, at fixed frequency, over the entire observed $\ell$-range. Specifically, at $\nu$=3.1\,mHz, the energy per mode drops by a factor of $\sim10$ between $\ell=150$ and $\ell=650$.  For flaring ARs, the mode energy variation rate is significantly different from the dormant ARs. $\alpha_1$ for flaring ARs increases with increasing $\ell$ except for few modes. On average, $\alpha_1$ of flaring ARs are smaller in magnitude than in dormant ARs indicating smaller mode energy in flaring ARs than in dormant ARs.

The above-mentioned difficulty may be resolved as follows:  It is to note that we have computed mode energy using data-cubes corresponding to the ARs of our sample during their maximum flaring periods. Hence, mode energy would include the net result of absorption due to the sunspots and amplification due to the flares. In ARs having large magnetic fields, i.e., large magnetic index ($B$) but lower flare activity, i.e., small flare index (FI), the effect due to mode absorption would dominate. On the other hand, when the flare energy is larger than the energy absorbed by sunspots, then the amplification effect due to flares would dominate. We suggest that the small slope in p-mode energy (Figure \ref{fig:freq_av_b0_area_mai}), not showing a significant decrease with increasing $B$, could be attributed to the increasingly larger flaring activity in the magnetically complex, flare productive ARs. However, it is not clear whether the variations in $A$ or in $\Gamma$ are contributing more to the changes in mode energy. In the following, we attempt to further analyse this issue.

\begin{figure*}[ht]
    \centering
        \includegraphics[width=0.9\textwidth,clip=,bb=12 7 458 501]{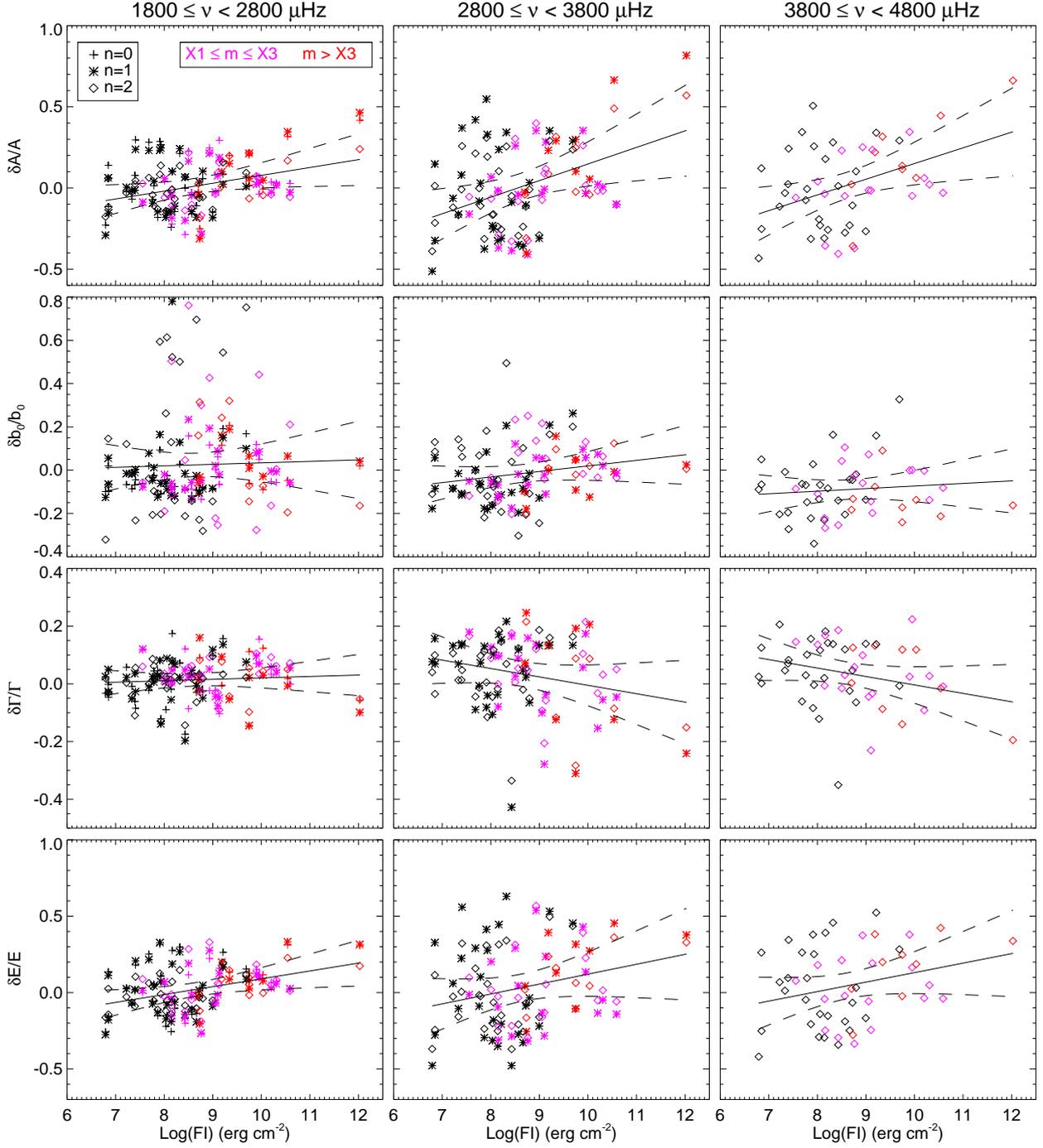}
    \caption{Frequency averages of the fractional difference in mode amplitude, background power, mode width and mode energy of flaring ARs, respectively from top to bottom, as function of flare activity index. Mode parameters are corrected for foreshortening, filling factor and magnetic activity. Red and magenta symbols correspond to those ARs which have flare(s) of magnitude (say, $m$) > X3 and X1$\leq m\leq X3$, respectively. Solid lines show the linear regression and dashed curves correspond to 95\% confidence level of the linear fit.}
    \label{fig:freq_av_modes_flr}
\end{figure*}

\begin{table}[ht]%
\centering
\caption{Linear regression ($Y=\alpha + \beta\,{\rm FI}$) and Pearson correlation coefficients ($R$) obtained from the frequency averaged fractional mode difference ($Y$) in three frequency bands and flare activity index (FI) of the flaring ARs. Note that the mode parameters are corrected for the foreshortening, filling factor and magnetic activity index.}
\begin{tabular}{|l|l|r|r|r|}
\hline
                       Mode  &      $\nu$ -range &$\alpha$ &  $\beta$ $\pm$          $\sigma$ &   $R$\\
\cline{2-5}
                 Parameters  & ($\mu Hz$)        &         &               $(\times10^{-2} {\rm erg}^{-1} {\rm cm}^2)$ &      \\
\hline
\hline
                                  &       1800 - 2799 &   -0.41 &     4.84 $\pm$              1.79 &  0.35\\
                     $\delta A/A$ &       2800 - 3799 &   -0.87 &    10.21 $\pm$              3.15 &  0.41\\
                                  &       3800 - 4800 &   -0.82 &     9.66 $\pm$              3.04 &  0.41\\
\hline
                                  &       1800 - 2799 &   -0.04 &     0.69 $\pm$              2.02 &  0.05\\
                 $\delta b_0/b_0$ &       2800 - 3799 &   -0.24 &     2.56 $\pm$              1.53 &  0.23\\
                                  &       3800 - 4800 &   -0.19 &     1.19 $\pm$              1.67 &  0.10\\
\hline
                                  &       1800 - 2799 &   -0.03 &     0.50 $\pm$              0.81 &  0.09\\
            $\delta\Gamma/\Gamma$ &       2800 - 3799 &    0.29 &    -2.91 $\pm$              1.63 & -0.24\\
                                  &       3800 - 4800 &    0.29 &    -2.93 $\pm$              1.46 & -0.27\\
\hline
                                  &       1800 - 2799 &   -0.42 &     5.15 $\pm$              1.69 &  0.39\\
                     $\delta E/E$ &       2800 - 3799 &   -0.53 &     6.49 $\pm$              3.35 &  0.26\\
                                  &       3800 - 4800 &   -0.49 &     6.22 $\pm$              3.16 &  0.27\\
\hline
\end{tabular}

\label{tab:mod-ratio-flr}
\end{table}

In order to study the major contribution to mode energy variation, we plot the coefficients of fractional amplitude variation and width variation for flaring and dormant ARs (see Figure~\ref{fig:ap_vs_wd_mai}). For dormant ARs, we find that the fractional increase in mode width is followed by the fractional decrease in  mode amplitude, supporting previous reports \citep{Rabello-Soares2008}. However, the increase in width is slightly faster than the decrease in amplitude, as evident from the negative slope of the linear regression. This shows that the fractional decrease in mode energy of dormant ARs, is caused by decrease in amplitude as well as in width, but the width increase contributes slightly more than the decrease in amplitude.

For the flaring ARs, the trend is quiet different from the dormant ARs. The increase in fractional difference in mode width is followed by the increase in fractional difference in mode amplitude, and resulting increase in fractional mode energy. This can be seen in Figures~\ref{fig:freq_mai_cf} and~\ref{fig:freq_mai_cf_nu} (bottom panel). The fractional increase in mode width is relatively larger than the fractional increase in mode amplitude. This shows that the net contribution to the variation in fractional mode energy is dominated by mode width.

\subsection{Changes in Mode Parameters with Flaring Activity}
\label{subsec:ModesFlr}

From the previous Section~\ref{subsec:ModPar-MAI}, we learned that the p-mode properties of flaring ARs are distinctly different from those of the dormant ARs. We have smaller mode suppression and width variation rates in flaring ARs than in dormant ARs. In order to study the flare related changes in mode parameters, we corrected the mode parameters, for foreshortening, filling factor and magnetic activity, of all the flaring ARs and corresponding QRs. From Section~\ref{subsec:ModPar-MAI} we found that the mode amplitude and width show linear relation with $B$ in agreement to the reports by \citet{Rabello-Soares2008}. We fit the foreshortening and filling factor corrected mode parameters of all the ARs and QRs as a function of $B$, $\mathcal{P}^{\rm c}_{n,\ell}(B)=\xi_0+\xi\,B$. Then every mode parameter of flaring ARs and corresponding QRs are corrected for magnetic activity index using the coefficients $\xi$ obtained from the fitting according to, $\mathcal{P}^{'\rm c}_{n,\ell}=\mathcal{P}^{\rm c}_{n,\ell}-\xi(B-B_{\rm min}) $, where, $B_{\rm min}$ is the minimum value of $B$ in the data sample. We repeated above analysis for every common multiplet ($n,\ell$) for mode parameters, $A$, $b_0$ and $\Gamma$. Then we computed the fractional difference in mode parameters between flaring ARs and corresponding QRs (e.g., for mode energy see Equation~\ref{eq:ModeEn-ratio}).

\begin{figure*}[ht]
    \centering
        \includegraphics[width=1.0\textwidth,clip=,bb=6 4 458 286]{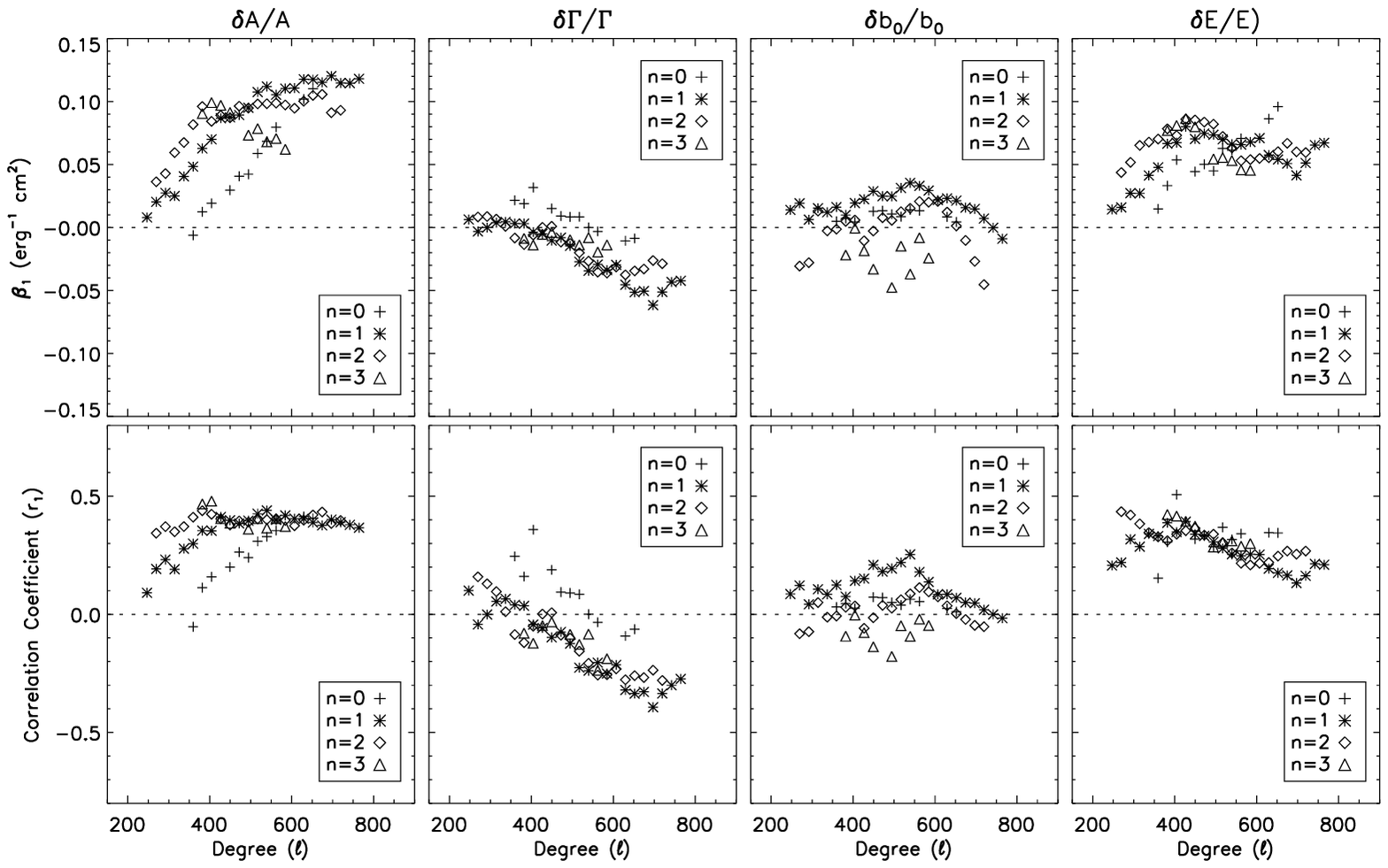}
    \caption{Coefficients of linear regression in FI (top row) and Pearson correlation (bottom row), for different mode parameters of flaring ARs as a function of harmonic degree.}
    \label{fig:freq_flr_cf}
\end{figure*}

\begin{figure*}[ht]
    \centering
        \includegraphics[width=1.0\textwidth,clip=,bb=6 4 458 286]{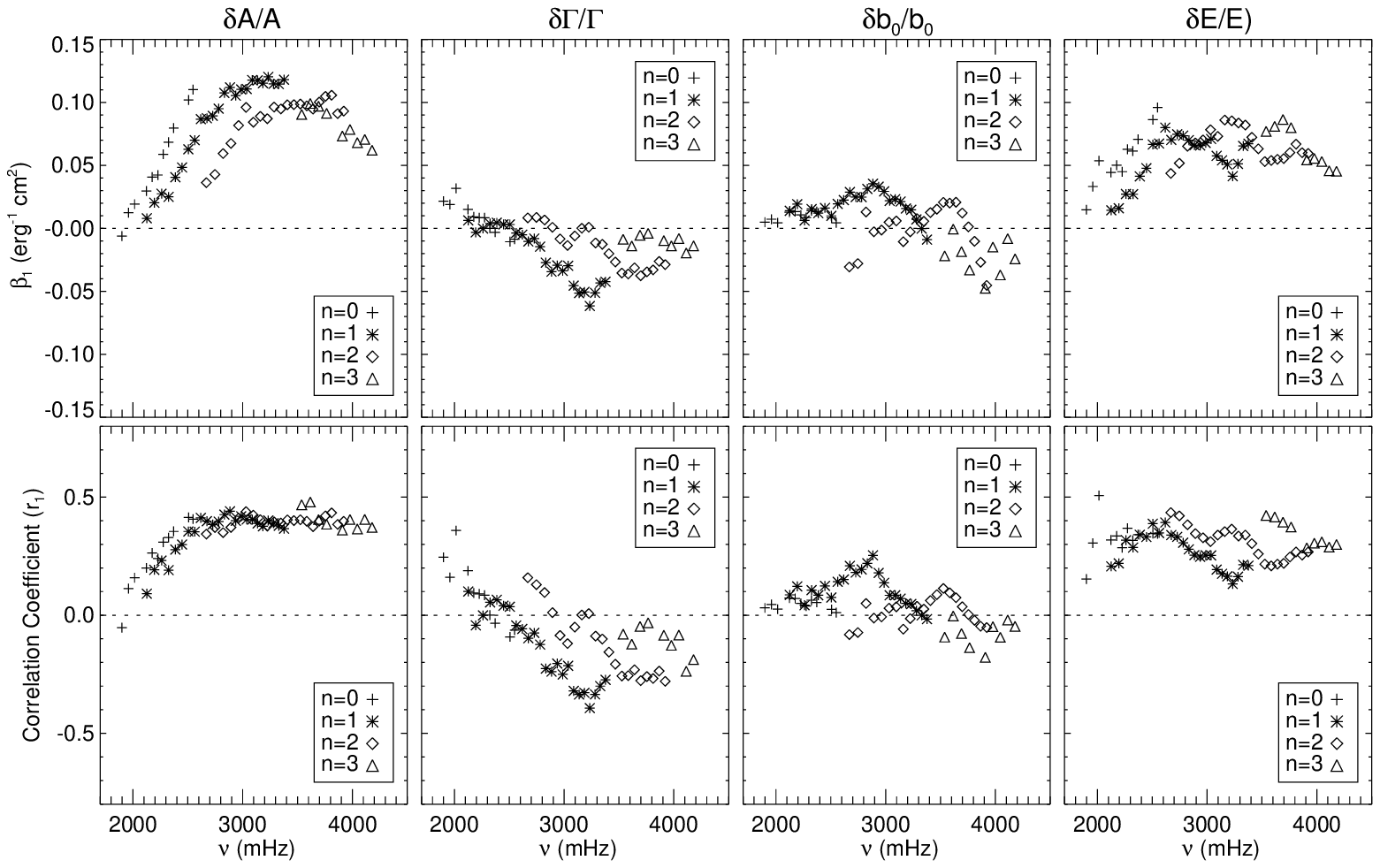}
    \caption{Similar to Figure~\ref{fig:freq_flr_cf} but as a function of frequency.}
    \label{fig:freq_flr_cf_nu}
\end{figure*}

Figure~\ref{fig:freq_av_modes_flr} shows the frequency average of the fractional mode difference of corrected mode parameters as function of flare activity index (FI). We computed the linear regression and correlation coefficients between the frequency averages of the fractional differences of mode parameters and flare indices which are listed in Table~\ref{tab:mod-ratio-flr}. Figures~\ref{fig:freq_flr_cf} and~\ref{fig:freq_flr_cf_nu} show the coefficients of linear regression and Pearson correlations between fractional difference in mode parameters and FI as a function of harmonic degree ($\ell$) and frequency ($\nu$), respectively. In the following, we analyse and discuss the various mode parameters.

\subsubsection{Mode Amplitude}

Figure~\ref{fig:freq_av_modes_flr} (top panel) shows that mode amplitude of flaring ARs increases with FI in all frequency bands. In the five minute band the increase in $\delta A/A$ is larger than the lower and higher frequency bands, as evident from the linear regression slope for the five minute band (see Table~\ref{tab:mod-ratio-flr}). Also the correlation between $\delta A/A$ and FI  is larger for the five minute and higher frequency bands than in the lower frequency band. The positive large slope shows flare associated enhancement in mode amplitude. Our results of flare induced amplification in mode amplitude supports earlier reports  \citep{Ambastha2003a, Maurya2009f}.

However, several data points with Log(FI)$\leq9$ are seen below $\delta A/A=0$, showing smaller amplitude in flaring ARs than corresponding QRs. Similar results have also been reported by \citet{Ambastha2003a} for some flaring ARs. Figures~\ref{fig:freq_flr_cf} and~\ref{fig:freq_flr_cf_nu} (left column) show the linear regression coefficient ($\beta_1$) as a function of harmonic degree ($\ell$) and frequency ($\nu$), respectively. The amplitude variation rate increases with $\ell$ for all the radial orders. It increases with frequency and becomes largest in the five minute band, then decreases. Also, we find significant correlation for degree $\ell>300$ and $\nu>2500$.

\subsubsection{Mode Width}

Figure~\ref{fig:freq_av_modes_flr} (third row from the top) shows the frequency average of fractional difference ($\delta \Gamma/\Gamma$) in mode width as a function of FI of the flaring ARs. $\delta \Gamma/\Gamma$ in the five minute and higher frequency bands decreases with increasing FI while it shows opposite trend in the lower frequency band. The relation between fractional difference in mode width and FI is also evident from the regression slope and correlation coefficients given in Table~\ref{tab:mod-ratio-flr}. There is a strong correlation between relative width and FI for five minute and higher frequency band but the correlation is poor at lower frequency band. \citet{Ambastha2003a} have found decrease in mode width during flares in some ARs, but no such signatures for some other flaring ARs in their data samples. \citet{Tripathy2008} have found that the CME-prone ARs having lower values of magnetic flux have smaller line width than the QRs.

Figures~\ref{fig:freq_flr_cf} and~\ref{fig:freq_flr_cf_nu} (second column from the left) show the linear regression coefficients $\beta_1$ for relative mode width as function of harmonic degree and frequency, respectively. The coefficient $\beta_1$ is very small for harmonic degree $\ell<450$ and decreases with increasing $\ell (>450)$. The correlation coefficients ($r$) also show similar trend as the parameter $\beta_1$. The coefficient $\beta_1$ decreases with frequency for different radial orders (Figure~\ref{fig:freq_flr_cf_nu}). But for the modes with radial order $n=1,\,2$, $\beta_1$ is smallest in the five minute band at different frequencies then it increases. The decrease in mode width with FI indicates increase in the lifetimes of modes.

\subsubsection{Background Power}

Figure~\ref{fig:freq_av_modes_flr} (second row from the top) shows that fractional difference ($\delta b_0/b_0$) in background power increases with FI in all three frequency bands; the slope increases with increasing frequency. The slope is larger in the five minute band than in lower and higher frequency bands similar to amplitude. The correlation coefficient is also larger in the five minute frequency band than in the lower and higher frequency bands as evident from Table~\ref{tab:mod-ratio-flr}. The positive slope at all frequency bands of the background power further suggests the flare induced amplification in mode power. But the correlation in the background power and FI are not significant. The background power variation with degree (Figure~\ref{fig:freq_mai_cf}) and frequency (Figure~\ref{fig:freq_mai_cf_nu}) show similar relations.

\begin{figure*}
    \centering
        \includegraphics[width=0.7\textwidth,clip=, bb=9 4 314 223]{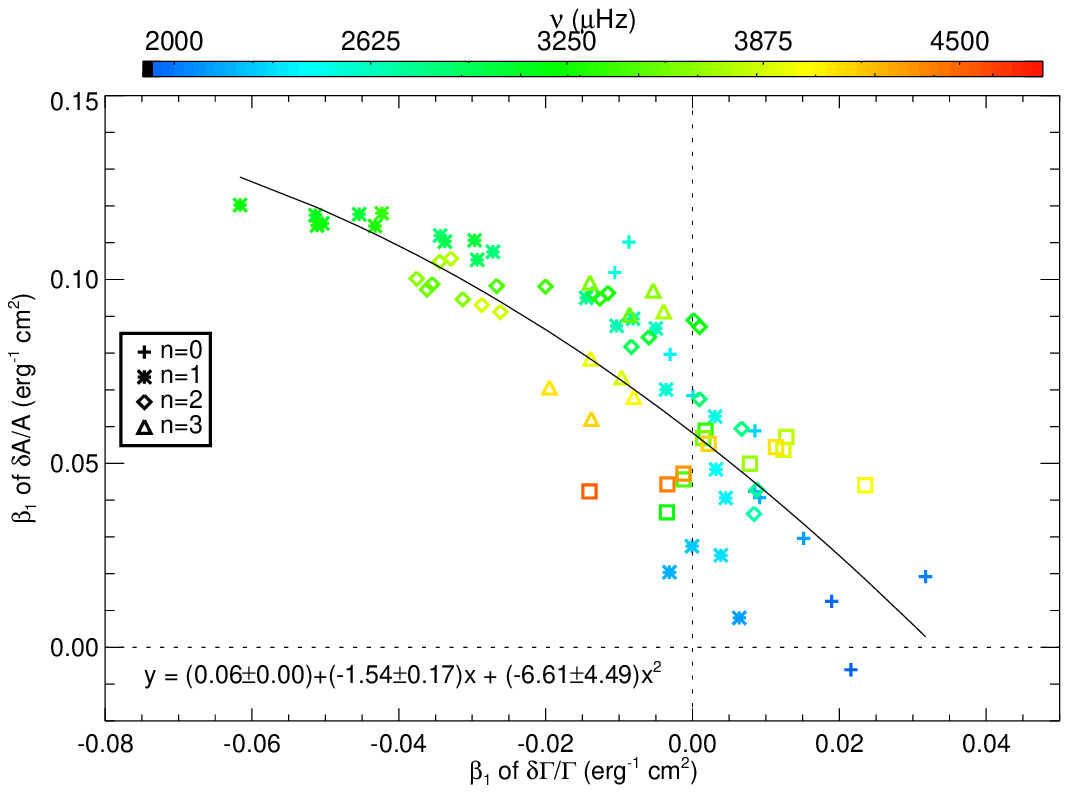}
    \caption{Variations in the rate of fractional difference in mode amplitude and width of flaring ARs. Solid line shows the second order polynomial fit; the fitted coefficients are given in the equation.}
    \label{fig:ap_vs_wd_flr}
\end{figure*}

\subsubsection{Mode Energy}

The frequency averaged fractional difference in mode energy ($\delta E/E$) in all three frequency bands increases with increasing flare activity index, see Figure~\ref{fig:freq_av_modes_flr}(bottom row). The rate of increase in $\delta E/E$ for the five minute frequency band is largest as evident from the magnitude of the linear regression slope which is smaller in the lower frequency band and in higher frequency band (see Table~\ref{tab:mod-ratio-flr}). This indicates that flare induced excitation is more significant in the five minute bands than in the lower and higher frequencies. The significant positive correlation coefficients between mode energy and FI further support above results.

Figures~\ref{fig:freq_flr_cf} and~\ref{fig:freq_flr_cf_nu} show that the mode energy variation rate with FI increase with increasing frequency and harmonic degree and peak in the degree range 400--600 and in the five minute frequency band for all radial orders, then decrease at higher degree and frequencies, respectively. In order to ascertain the contribution of amplitude and width in the energy variation, we plot the width variation rate vs. amplitude variation rate in Figure~\ref{fig:ap_vs_wd_flr}.

Figure~\ref{fig:ap_vs_wd_flr} shows that mode amplitude (mode width) variation rate is positive (negative) for most of the modes, and increase in mode width is followed by rapid decrease in mode amplitude. This shows the larger contribution of mode amplitude to the variation of mode energy than the mode width. In the five minute frequency band, there is a large positive amplitude variation rate and relatively small negative mode width rate. However, for a few modes in the lower and higher frequency bands the positive amplitude variation rate decreases very rapidly and positive mode width variation rate occurs.

\section{Summary and Conclusions}
 \label{sec:SumConclsn}

We studied high degree p-mode properties of a sample of several flaring and dormant ARs, and associated QRs, observed during the solar cycles 23 and 24 using the ring-diagram technique, under the presumption of plane waves, and their association with magnetic and flare activities. The changes in p-mode parameters are combined effects of filling factors, foreshortening, magnetic and flare activities, and measurement uncertainties.

The p-mode amplitude ($A$) and background power ($b_0$) of ARs are found to be decreasing with their angular distances from the disc centre while width increases slowly. The effects of foreshortening on the mode amplitude and width are consistent with reports by \citet{Howe2004}. The decrease in mode amplitude $A$ with distance is due to the fact that as we go away from the disc centre, we measure only the cosine component of the vertical displacement. Also, foreshortening causes a decrease in spatial resolution of Dopplergrams as we observe increasingly toward the limb. This reduces the spatial resolution determined on the Sun in the centre-to-limb direction, and hence leads to systematic observational errors.

The second largest effects on p-mode parameters are due to filling factor. We found that the mode amplitude increases with increasing filling factor while mode width and background power show opposite trend. Similar results are reported earlier for the global p-mode amplitude and width, for example by \citet{Komm2000}. They reported largest increase in mode width and reduction in amplitude with filling factor when its values are lower. Such changes in mode parameters may be caused by increase in signal samples in data cubes. However, we find that for a few modes in five-minute and in higher frequencies, mode amplitudes do not increase significantly with filling factor. The effect of filling factor decreases with increasing harmonic degree $\ell$. In order to study the relation of mode parameters with magnetic and flare activities, we corrected mode parameters of all the ARs and QRs for foreshortening.

We found that the mode amplitude in ARs are considerably smaller than in QRs. In the dormant ARs, the mode amplitude decreases with increasing $B$. There is a larger reduction in the five minute band than in the lower and higher frequency bands. The reduction in mode amplitude of ARs has been reported earlier  by several researchers \citep{Braun1987, Braun1990, Rajaguru2001, Mathew2008}.  \citet{Goode1996} have reported the suppression of acoustic flux and p-mode power even in a weak magnetic field QR. However, a precise mechanism of the absorption of energy is not yet established. The possible mechanisms for mode power reduction in ARs are: (i)  Absorption of p-modes within sunspots \citep{Braun1988}. It is believed that the sunspot magnetic fields play crucial role to transform some part of the acoustic energy to pure magnetic energy, e.g., Alfv\'en type \citep{Crouch2003, Cally2003, Crouch2005}, which may be transported to the upper atmosphere of the Sun \citep[][and references therein]{Marsh2006}.  (ii)  The efficiency of p-mode excitation by turbulent convection \citep{Goldreich1977, Goldreich1988, Goldreich1990} might be reduced in a magnetic field, owing to the nature of magnetoconvection \citep{Hughes1988}. (iii)  Wilson depression in sunspot, height of formation of spectral lines, used to measure the Doppler shift, causes additional phase shifts to the velocity measurements, as compared to non-magnetic regions, (iv) Modification of the surface values of the p-mode eigenfunctions \citep{Hindman1997}, (v) Resonant absorption \citep{Hollweg1988}, (vi) Mode mixing \citep{DSilva1994}. The energy in an incoming mode, at any horizontal wavenumber, get dispersed into a wide range of wave numbers. (vii) Excitation of tube waves through p-mode buffeting \citep{Bogdan1996, Hindman2008}, and (viii) Inhomogeneity-enhanced thermal damping \citep{Riutova1984}. \citet{Jain1996} show that the horizontal magnetic field can lower the upper turning point and change the skin depth for a simple plane-parallel adiabatically stratified polytrope. In addition to power suppression, they also found that magnetic field alters phase of p-modes.

The inclination of field lines from the vertical can affect the amount of acoustic power absorption \citep{Cally2003} by conversion of acoustic to slow magneto-acoustic waves. The observational confirmation about the field inclination related variations in mode power have been reported earlier from the time-distance \citep{Zhao2006} and acoustic-holography \citep{Schunker2005, Schunker2008} analyses. In ring-diagram analysis, we find these effects averaged over larger area.

The mode width in ARs are generally smaller than in corresponding QRs and increases with magnetic field. This may be caused by large damping of p-modes in the strong magnetic field areas. Also the width increases with frequency and becomes largest in the five minute bands then decreases. But the relation in flaring ARs is poor than in dormant ARs. This may be caused by flare induced changes in mode parameters. For stochastically excited modes, a broadening in mode width shows a reduced lifetime or increased damping of the modes in regions of high magnetic activity indices.

A possible mechanism by which magnetic activity can influence mode widths is excitation of oscillations in flux tubes as suggested by \citep{Bogdan1996, Hasan1997}. They suggested that the flux tubes lead to a balance between energy input from p-modes and losses through radiative damping and leakage from flux tube boundaries. The excitation of resonant oscillation in flux tubes in unstratified atmosphere is studied by \citet{Chitre1991}, and \citet{Ryutova1993, Ryutova1993a}. Resonant coupling with MHD waves \citep{Pinter1999} might also contribute the damping of p-modes, as well as scattering of p-modes by flux tube \citep{Keppens1994, Bogdan1987}. Thus, when $B$ increases, p-modes are increasingly damped by the interactions with the increasing number of flux tubes. \citet{Gascoyne2009} have shown that the suppression of sound speed and pressure within the flux tube region is not the only factor to consider in the scattering of p-modes. There is a direct effect of the magnetic fields caused by the flaring of field lines on phase shifts.

The combined effects of the mode amplification and width variations appear in mode energy variations. We find that the average mode energy in flaring ARs are smallest at all frequencies than in dormant ARs and QRs. The rate of decrease of mode energy in dormant ARs are largest around $\ell=450$ while the maximum decrease rate for flaring ARs shifted towards lower $\ell$. More interestingly, we found that the increase in width variation rate of dormant (flaring) ARs is followed by decrease (increase) in mode amplitude variation rate. The first one for the dormant ARs confirms the earlier reports by \citet{Rabello-Soares2008}. The second one for the flaring ARs attributes to the flare associated changes in the oscillations characteristics of p-modes.

The decrease in mode energy implies a reduction in the amount of acoustic energy pumped into the modes by turbulent convection. This agree with the assumption that the presence of strong magnetic fields suppresses motion in a turbulent medium, which is known to occur for several solar surface activities. For turbulent excitation and damping, the energy of a single oscillation mode is expected to depend mainly on mode frequency (at least for modes whose angular wavenumber is well below that of the excitation turbulence). The implication of energy depending only on frequency is that the surface velocity power of a single oscillation mode should increase with angular degree at fixed $\nu$. Theoretically, \citet{Jain2009a} have shown that mode absorption increases rapidly with frequency reaching a maximum or saturation value near 4\,mHz and decreases at higher frequencies $\nu>4$\,mHz. They investigated the p-mode absorption by fibril magnetic field. They suggested that the p-modes excite tube waves, through mechanical buffeting, on the magnetic fibrils in the form of longitudinal sausage waves and transverse kink waves. The tube waves propagate up and down the magnetic fibrils and out of the p-mode cavity, thereby removing energy from the incident acoustic waves.

The background power of the dormant ARs are found to be smaller than in corresponding QRs for radial order $n\leq2$. But for the flaring ARs, the background power variations seem to be function of harmonic degree and frequency ($\nu$). It decreases with increasing $\nu$ and becomes smallest at specific frequency in the five minute frequency band the increases. We believe that such changes are attributed to the flare associated charges in the oscillation modes. The background power of flaring and dormant ARs are found to decrease with magnetic activity index which reinforce the idea that high magnetic fields could hinder convection \citep{Biermann1941, Chandrasekhar1961a}, which is the source of solar noise.

We believe that magnetic field induced activities in the surface and higher layers may play a role in the excitation of oscillation modes in the ARs. In order to study the effects of flares on p-mode parameters, we employed mode corrections for foreshortening, filling factor and magnetic activity. We find that the mode parameters show significant correlation with flare activity index. The p-mode amplitude increases with flare activity index and shows stronger amplification in the five minute band. The increases in background power with flare activity index further supports the flare associated excitation in p-modes \citep{Wolff1972}. More interestingly, our statistical study shows association of p-mode energy with flares, supporting the expected mode excitation by flares. The mode width is found to decrease with flare activity index indicating increase in lifetime of modes or mode damping. The combined effects of mode amplitude and width appeared in the mode energy. The mode energy increases with increasing flare activity index and more largely so in the five minute band. Our study shows that the increase in the mode energy is mostly contributed by the increase in mode amplitude rather than the increase in width or damping.

This suggests that the energetic solar flares may indeed produce effects in the solar interior below ARs although the major process of flare energy release occurs in the external solar atmosphere. However, the exact mechanism of the energy transfer towards solar photosphere and flare induced excitation in mode power is not yet known. The detailed mechanism of the transfer of energy from the corona to the photosphere is not well understood. \citet{Wolff1972} probably first proposed the mechanism of momentum transfer toward the photosphere from the flare energy release site to describe the possible impulse for the excitation of the p-modes. Recently, \citet{Hudson2012} proposed similar mechanism of momentum transfer to explain flare induced excitation of the seismic waves. They listed four basic mechanisms proposed earlier: (i) hydrodynamic shock-wave originating in the chromosphere \citep{Kostiuk1975, Kosovichev1998}, (ii) Lorentz force from the magnetic transients \citep{Anwar1993, Kosovichev2001, Sudol2005, Hudson2008}, (iii) photospheric back-warming \citep{Machado1989, Martinez-Oliveros2008}, and (iv) ``McClymont magnetic jerk'' \citep{Hudson2008}. Work done by Lorentz forces in the back-reaction could supply enough energy to explain observations of flare-driven seismic waves \citep{Hudson2008}. The requirement for momentum conservation can in principle help to distinguish among these plausible mechanisms \citep{Hudson2012}. \citet{Hudson2008} introduced the idea of the coupling of flare energy into a seismic wave, namely the ``McClymont magnetic jerk'', produced during the impulsive phase of acoustically active flares.

\citet{Maurya2009a} and \citet{Maurya2012} have reported large X-class white-light flares in seismically active ARs. \citet{Pedram2012} have  also studied the Hard X-Ray characteristics of seismically active and quiet white-light flares. They found that the acoustically active flares are associated with a larger and more impulsive deposition of electron energy.  However, they argued that this does not always correspond to a higher white-light contrast. Unfortunately, the observations of white-light flares are rare. We expect that the X-ray flux does not seem to be an adequate measure of how much energy a flare deposits in the photosphere. It is mainly because of the extent of soft X-ray emission produced at higher coronal heights. A smaller X-class but larger H$\alpha$-class or a white-light flare may be more important for affecting the solar acoustic oscillation modes.

We find the the flare index (FI), as calculated in this study, is not a good indicator for the helioseismic effects of a flare on the photosphere and on the oscillations characteristics as it is primarily based on the X-ray flux alone. Also, a larger value of FI may be result of several relatively longer duration flare of smaller magnitude integrated over the time internal of the ring data cube. These small magnitude flare that contribute to the larger FI may not be able to significantly affect the oscillations characteristics. On the other had it is likely that a short duration, energetic and impulsive flare of large magnitude is more effective but may have only a smaller FI.

The magnetic activity index (MAI or $B$) used in this analysis are obtained from the line-of-sight components of the full magnetic field strength. The mode power absorption also depends upon the inclination of field lines from the vertical as stated above. This may give rise to some systematic errors in the analysis of the relationship of mode parameters and magnetic activity. Different magnetic field configurations in ARs may give  different systematic errors. It is difficult to account for the order of the systematic errors but continuous observations of vector magnetic fields can be used to analyse the errors.

This study supports the earlier results by \citet{Ambastha2003a} and \citet{Maurya2009f} that large flares are able to significantly amplify the high degree p-modes, over and above the mode absorption by strong magnetic fields of ARs. The associated p-mode parameters are also affected by the flare-induced changes, thereby affecting the sub-surface properties that are derived using the computed p-modes. Therefore, we suggest that adequate care should be taken in describing the sub-surface properties of ARs, specially while using the photospheric p-modes. We plan to further analyse the frequency shift, sub-photospheric flow, sound speed, etc. in these active regions in our future studies.

\begin{acknowledgements}
This work utilizes data obtained by the GONG program  operated by AURA, Inc. and managed by the National Solar Observatory under a cooperative agreement with the National Science Foundation, U.S.A. The ring-diagram analysis is performed using the GONG pipeline. The integrated X-ray flux data was obtained from GOES which is operated by National Oceanic and Atmospheric Administration, U.S.A. The solar activity information provided by the solar monitors web pages help us to select the flaring and dormant active regions. The authors would like to thanks H.M. Antia and F.Hill for their useful discussion and suggestions. Authors, R.A.M and J.C. acknowledge support by the National Research Foundation of Korea (2011-0028102 and NRF-2012R1A2A1A03670387).
\end{acknowledgements}

\end{document}